\pdfoutput=1
\documentclass[pdflatex,sn-nature]{sn-jnl}

\usepackage{graphicx}%
\usepackage{amsmath,amssymb,amsfonts}%
\usepackage{xcolor}%
\usepackage{textcomp}%

\setcitestyle{super,open={},close={}}

\ifdefined\iflatexml
  \usepackage{authblk}
  \newcommand{\corrauthor}{\author}
  \providecommand{\fnm}[1]{#1}
  \providecommand{\sur}[1]{#1}
  \providecommand{\orgdiv}[1]{#1}
  \providecommand{\orgname}[1]{#1}
  \providecommand{\orgaddress}[1]{#1}
  \providecommand{\city}[1]{#1}
  \providecommand{\postcode}[1]{#1}
  \providecommand{\country}[1]{#1}
\else
  \newcommand{\corrauthor}{\author*}
\fi

\begin{document}

\title[Origins of Universal Machine Learning Force-Field Errors]{Origins of Universal Machine Learning Force-Field Errors in Multicomponent Materials}

\corrauthor[1,2]{\fnm{Hongwei} \sur{Du}}\email{duhongwei@zgci.ac.cn}

\author[1,2]{\fnm{Dingyang} \sur{Lv}}

\author[1,2]{\fnm{Baole} \sur{Wei}}

\author[3]{\fnm{Yu} \sur{Ren}}

\author[1,2]{\fnm{Feng} \sur{Yu}}

\author[1,2]{\fnm{Xin} \sur{He}}

\author[6]{\fnm{Bonan} \sur{Zhu}}

\author[1,2]{\fnm{Jiahui} \sur{Liu}}

\author[1,2]{\fnm{Yongda} \sur{Huang}}

\author[1,2]{\fnm{Yongheng} \sur{Li}}

\author[7]{\fnm{Jianjun} \sur{Liu}}

\author[5]{\fnm{Siqi} \sur{Shi}}

\author[4]{\fnm{Hong} \sur{Wang}}

\corrauthor[1,2,3]{\fnm{Ziheng} \sur{Lu}}\email{luziheng@zgci.ac.cn}

\affil[1]{\orgname{Zhongguancun Academy}, \orgaddress{\city{Beijing} \postcode{100094}, \country{China}}}

\affil[2]{\orgname{Zhongguancun Institute of Artificial Intelligence}, \orgaddress{\city{Beijing} \postcode{100094}, \country{China}}}

\affil[3]{\orgname{Kairos Materials}, \orgaddress{\city{Beijing}, \postcode{100094}, \country{China}}}

\affil[4]{\orgdiv{School of Materials Science and Engineering}, \orgname{Shanghai Jiao Tong University}, \orgaddress{\city{Shanghai} \postcode{200240}, \country{China}}}

\affil[5]{\orgdiv{State Key Laboratory of Materials for Advanced Nuclear Energy \& School of Materials Science and Engineering}, \orgname{Shanghai University}, \orgaddress{\city{Shanghai} \postcode{200444}, \country{China}}}

\affil[6]{\orgdiv{School of Aerospace Engineering}, \orgname{Beijing Institute of Technology}, \orgaddress{\city{Beijing} \postcode{100081}, \country{China}}}

\affil[7]{\orgdiv{Shanghai Institute of Ceramics}, \orgname{Chinese Academy of Sciences}, \orgaddress{\city{Shanghai} \postcode{201899}, \country{China}}}

\abstract{Anion mixing and high-entropy compositional design are widely used to tune materials properties, yet the ability of universal machine learning force fields to generalize to the resulting local environments remains insufficiently assessed. Existing training datasets expand chemical and structural space through relaxation trajectories, stability-oriented materials searches and non-equilibrium sampling, with sampling objectives that differ from property-driven compositional design. Here we construct a benchmark of 7,599 multicomponent computational configurations inspired by high-entropy design, elemental substitution and anion mixing. We evaluate eleven pretrained models against density functional theory for energies, forces and stresses, and extend the assessment to elastic, vibrational and adsorption-related properties. We investigate force errors first through training-reference coverage, then through local geometric heterogeneity, distance directionality and elemental response. Distances to training-reference environments reveal a qualitative association between coverage differences and increasing errors, while substantial error variation remains at similar distances. Element-aligned Voronoi distributions show that higher-error configurations extend towards greater geometric heterogeneity. However, OMat24 already provides broad coverage of these environments, motivating a closer examination of their interatomic-distance variations and chemical context. Separating distance deviations into compression and extension relative to the training-reference pair median reveals a pronounced asymmetry: errors remain low near the median, rise steeply on the compression side and increase more weakly on the extension side. After matching element pairs and absolute distance deviations, compression-side force errors are 1.81--1.95 times extension-side errors. Model-predicted pairwise interaction curves also exhibit greater curvature under compression, consistent with different interaction sensitivities in the compressed and extended regions. Elemental response helps explain how fitting difficulty varies with chemical environment. Fitting difficulty in independent elemental systems correlates with electronic band-energy responses to atomic displacements and Fermi-level shifts, and its overall trend persists in element-resolved errors in multicomponent systems. In parameter-matched comparisons, second- and fourth-order spherical-harmonic representations lower test force errors for 38 and 40 of 43 elements, respectively, while differences in elemental difficulty remain. The analysis connects qualitative coverage differences to specific geometric and electronic responses that correlate with force errors. These findings inform force-field selection for experimental compositional design and identify targets for training-data sampling and model representations.}

\maketitle

\section{Introduction}\label{sec:introduction}

Compositional design tunes materials properties through elemental substitution, anion mixing and high-entropy strategies\cite{yeh2004,rost2015,vandermaas2023}. Mixed chloride--bromide solid electrolytes demonstrate the effects of anion substitution on ionic conductivity and electrochemical stability\cite{vandermaas2023}, while multi-principal-element alloys and entropy-stabilized oxides substantially expand the local chemical environments accessible within a lattice\cite{yeh2004,rost2015}. Accurately predicting the resulting property changes requires interatomic interaction models that describe bonding across different combinations of elements and geometries. Density functional theory (DFT) provides an established computational reference, but its cost limits extensive sampling of multicomponent configuration space. Universal machine learning force fields enable larger-scale calculations by learning interatomic interactions from large first-principles datasets\cite{batatia2025,deng2023,barrosoluque2026}. For experimental compositional design, their value ultimately depends on whether these learned interactions transfer to the chemical and structural environments produced by mixing and substitution.

Universal force fields have developed through larger training datasets, changes in model architecture and broader evaluation. The data and benchmarking efforts provide the context for testing transfer to multicomponent environments. MPtrj samples intermediate and relaxed configurations from Materials Project first-principles relaxation trajectories\cite{deng2023}. Alexandria combines machine-learning-assisted searches for stable compounds with first-principles calculations to explore composition and crystal-structure space\cite{schmidt2023}. Starting from Alexandria structures, OMat24 broadens configuration sampling through Boltzmann sampling of randomly perturbed configurations, ab initio molecular dynamics and relaxation of perturbed structures\cite{barrosoluque2026}. These datasets include both relaxed and non-equilibrium configurations and substantially extend chemical and structural coverage. On the evaluation side, controlled benchmarks compare model accuracy and computational cost on shared data\cite{zuo2020}. Matbench Discovery connects energy prediction and structural relaxation to the identification of stable materials\cite{riebesell2025}. Forces Are Not Enough shows that low static errors do not guarantee simulation quality\cite{fu2023}. Phonon and solid-electrolyte benchmarks examine vibrational, mechanical and transport properties\cite{loew2025,du2025}. CHIPS-FF and LAMBench extend evaluation to computational convergence, cost and applicability across tasks and materials domains\cite{wines2025,peng2026}. UniFFBench further introduces experimental mineral structures and measured properties to examine discrepancies beyond agreement with DFT labels\cite{mannan2026}. These benchmarks test whether prediction accuracy carries over to practical materials calculations.

Studies of error mechanisms have further identified environmental and physical sources of prediction error. Surface benchmarks relate errors to the departure of test environments from the training domain\cite{focassio2025}. Studies of systematic softening find widespread underestimation of energies and forces in surfaces, defects, solid solutions, phonons and high-energy configurations\cite{deng2025}. Equation-of-state tests and structural searches in elemental systems reveal potential-energy-surface errors that near-equilibrium tests can miss\cite{tahmasbi2026}. Metal-43 further relates elemental fitting difficulty to the response of electronic band energy to atomic displacements and Fermi-level shifts\cite{geng2026}. These studies relate prediction errors to training environments and the interactions that models must reproduce.

For multicomponent configurations inspired by experimental compositional design, however, how environmental coverage, geometric features and elemental response jointly affect model accuracy remains insufficiently understood. Relaxation datasets, stability-oriented structure searches and non-equilibrium perturbation sampling serve different objectives from property-driven anion mixing and high-entropy design. Broad elemental coverage therefore does not ensure adequate sampling of the local combinations generated by these strategies. Explaining the magnitude and physical character of these errors requires a closer examination of local geometry and chemistry. Geometric heterogeneity describes the spread of distances within a coordination shell, while signed distance deviations distinguish compression from extension and their different interaction responses. Element-dependent fitting difficulty adds another influence, whose persistence in mixed chemical environments requires examination. Addressing these questions requires a common multicomponent benchmark in which reference-environment coverage, geometric heterogeneity, distance directionality and elemental response are examined in sequence.

Here we construct a benchmark of 7,599 computational configurations inspired by high-entropy compositional design, elemental substitution and anion mixing, using explicit compositional selection criteria and comparisons with training-reference data. We evaluate eleven pretrained models against common DFT energy, force and stress labels and extend the assessment to elastic, vibrational and adsorption-related properties. We then analyse force errors through environmental coverage, geometric heterogeneity, distance directionality and elemental response. Distances to training-reference environments describe coverage differences and their qualitative association with error. Element-aligned comparisons show that higher-error configurations extend towards greater Voronoi geometric heterogeneity, while OMat24 retains broad coverage of these environments. This overlap motivates a finer analysis of distance directionality and elemental response. After matching element pairs and distance-deviation magnitudes, force errors on the compression side are 1.81--1.95 times those on the extension side. Model-predicted pairwise interactions show greater curvature under compression, providing a qualitative physical context for this asymmetry. Electronic responses to atomic displacements and Fermi-level shifts help explain elemental fitting-difficulty trends that persist in multicomponent systems. Matched MACE comparisons further test the influence of angular representation on elemental force fitting. A structural error predictor complements these analyses by quantifying the additional information supplied by local chemistry and geometry beyond short-distance descriptors. The results identify environments and interaction responses that can guide training-data sampling and model representations for experimental compositional design.

\section{Results}\label{sec:results}

\subsection{Study overview}

\begin{figure}[!t]
\centering
\includegraphics[width=\textwidth]{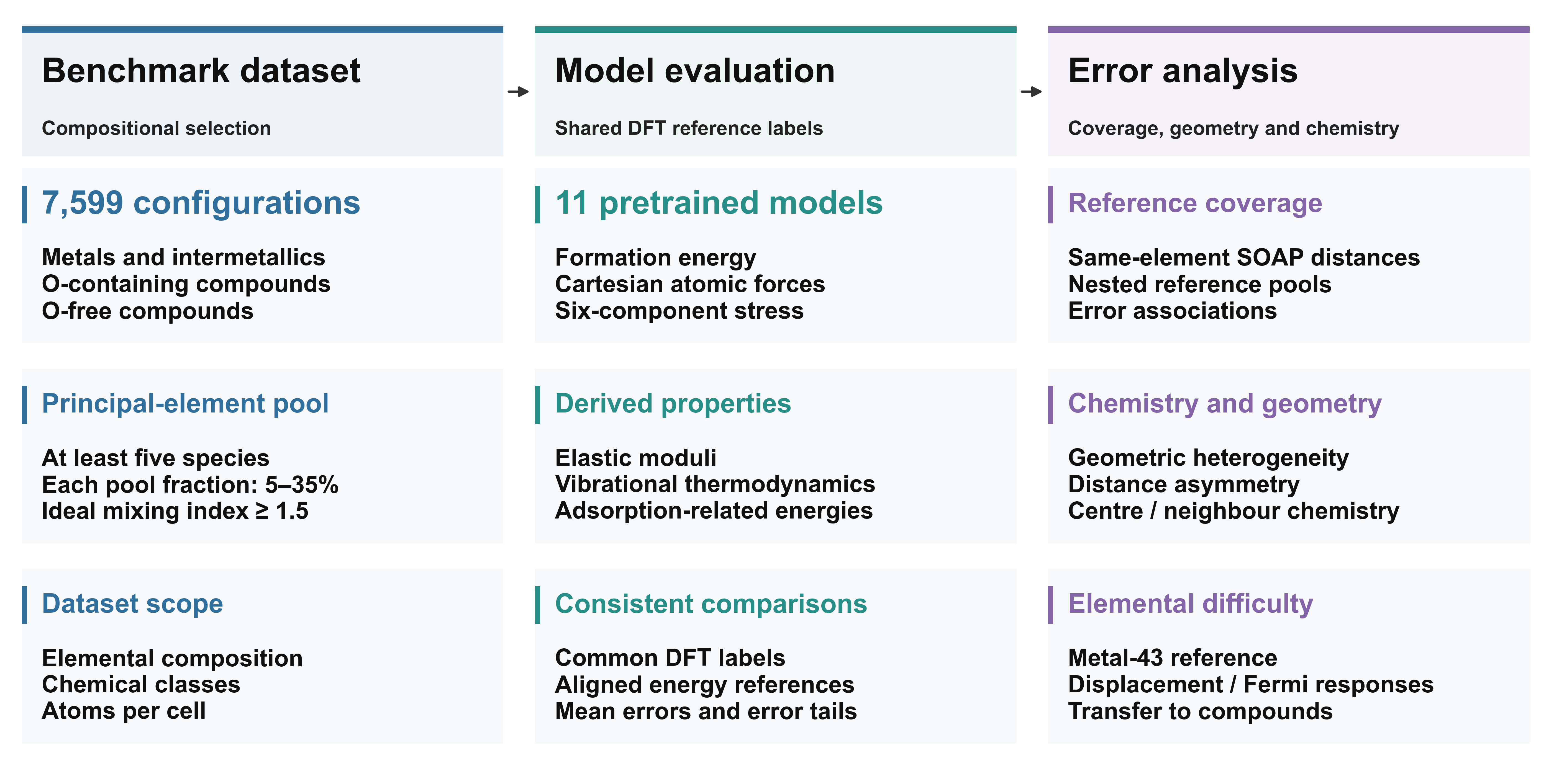}
\caption{\textbf{Overview of the benchmark, model evaluation and structural error analysis.} Left, 7,599 configurations are selected using a principal-element pool with at least five species, pool fractions of 5--35\% and an ideal mixing index $H_{\mathrm{pool}} \geq 1.5$. Centre, eleven pretrained force fields are evaluated against common DFT reference energies, forces and stresses with aligned reference conventions. The same eleven models are also compared on derived properties. Right, structural error analysis examines training-reference coverage, followed by local chemical and geometric heterogeneity, asymmetric responses to interatomic-distance compression and extension, and elemental fitting difficulty assessed using an independent elemental-metal reference.}\label{fig:1}
\end{figure}

We test how pretrained force fields transfer to multicomponent configurations motivated by experimental compositional design (Fig.~\ref{fig:1}). We define a benchmark of 7,599 compositionally qualified configurations and evaluate eleven models using common structures, labels and error definitions. We extend the comparison to elastic, vibrational and adsorption-related properties, then examine force errors from environmental coverage to local geometric heterogeneity, directional interatomic-distance variations and elemental response. An atom-based error predictor provides a quantitative overview of the structural and chemical information associated with error. The subsequent analyses first compare element-aligned Voronoi heterogeneity, then resolve distance compression and extension, and finally examine elemental electronic response and angular representation.

\subsection{Multicomponent configuration dataset}

\begin{figure}[!t]
\centering
\includegraphics[width=\textwidth]{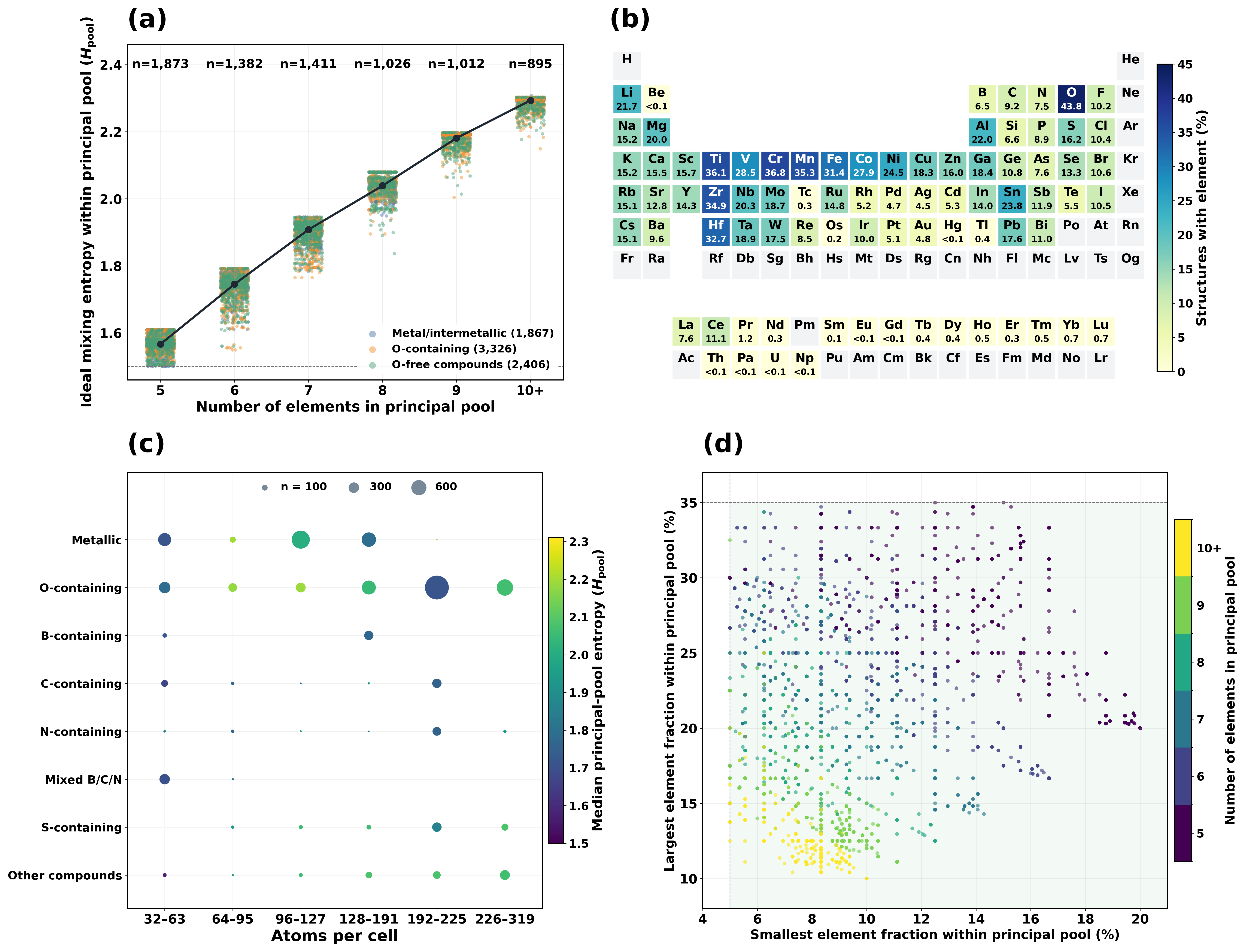}
\caption{\textbf{Composition and size of the benchmark configurations.} a, Dimensionless principal-pool ideal mixing index $H_{\mathrm{pool}} = -\sum_{e} x_{e} \ln(x_{e})$ versus the number of elements in the principal pool. The pool is specified by the structure-design record. Here $x_{e} = n_{e}/\sum_{e} n_{e}$ is the atomic fraction of element $e$ within that pool, $\sum_{e} x_{e} = 1$, and $\ln$ denotes the natural logarithm. The groups with 5, 6, 7, 8, 9 and at least 10 elements contain 1,873, 1,382, 1,411, 1,026, 1,012 and 895 configurations, respectively. Colours distinguish metal/intermetallic, oxygen-containing and oxygen-free compound configurations, with counts of 1,867, 3,326 and 2,406. The black line connects group medians, and the horizontal dashed line marks $H_{\mathrm{pool}} = 1.5$. b, Elemental distribution of the benchmark. Cell colours and values indicate the percentage of configurations containing each element. c, Joint distribution of chemical class and atoms per cell. Bubble area represents the number of configurations, and colour represents the median principal-pool mixing index in each group. d, Smallest versus largest elemental fraction within the principal pool. Colour indicates the number of elements in the pool, with counts of 10 or more grouped as 10+. Dashed lines mark the 5\% and 35\% fraction thresholds.}\label{fig:2}
\end{figure}

The benchmark contains 7,599 configurations with complete energy, force and stress evaluations for all eleven models. Principal-element substitution and anion mixing generate the multicomponent environments examined here. Figure~\ref{fig:2} summarizes their composition and cell size. The principal mixing pool specified in each structure-design record contains at least five species, each contributing 5--35\% of its atoms. Its dimensionless ideal mixing index is $H_{\mathrm{pool}} = -\sum_{e} x_{e} \ln(x_{e})$, where $x_{e}$ is the pool atomic fraction of element $e$, $\sum_{e} x_{e} = 1$, and $\ln$ denotes the natural logarithm. All selected configurations satisfy $H_{\mathrm{pool}} \geq 1.5$. In Fig.~\ref{fig:2}a, the index rises with the number of principal elements, while its spread at each count reflects differences in their proportions. Equal proportions give the upper bound $H_{\mathrm{pool}} = \ln(m)$ for $m$ pool elements. Figure~\ref{fig:2}d complements this view: as the pool contains more elements, the smallest and largest fractions draw closer together, forming a more balanced composition range. Together, these panels show how the selection combines multiple principal species with substantial contributions from each species.

\begin{figure}[!t]
\centering
\includegraphics[width=0.92\textwidth]{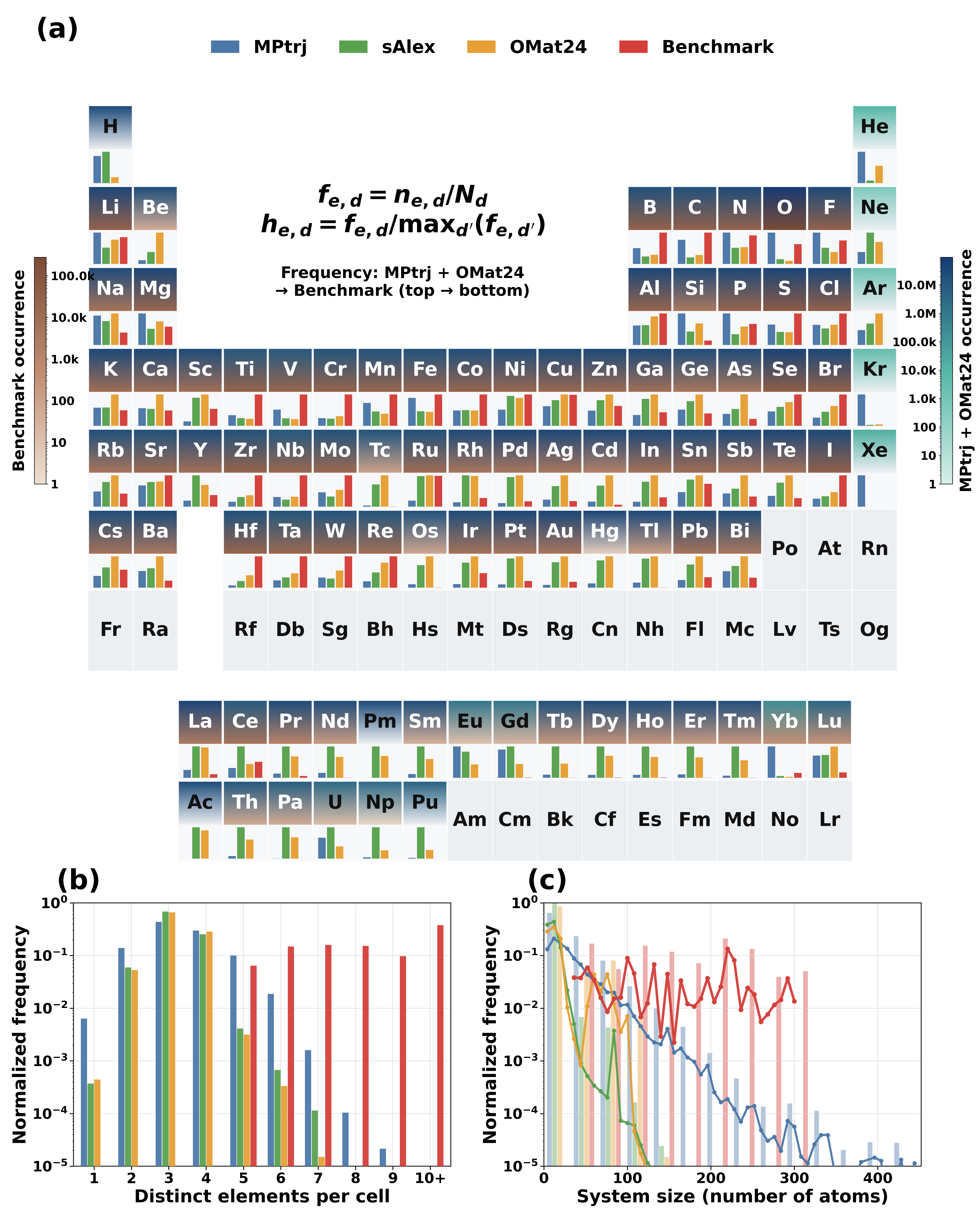}
\caption{\textbf{Training-reference coverage of elements and structural complexity.} a, Elemental occurrence in MPtrj, sAlex, OMat24 and the benchmark. Bars show the fraction of structures containing each element, normalized by that element's largest fraction among the datasets. Tile shading compares training and benchmark occurrence using the indicated scales. b, Normalized distributions of the number of distinct chemical elements per cell. For all four datasets, cells containing ten or more distinct elements are combined in the 10+ category. Frequencies are normalized over each full dataset, including all 7,599 benchmark configurations. c, Normalized distributions of the number of atoms per cell. The vertical axes in b and c are logarithmic. These all-formula element counts differ from the principal-pool counts used for compositional selection.}\label{fig:3}
\end{figure}

The chemical and size distributions show how this design is realized across material families. The benchmark comprises 1,867 metal or intermetallic configurations, 3,326 oxygen-containing configurations and 2,406 oxygen-free compounds. Figure~\ref{fig:2}b shows frequent occurrence of transition metals alongside oxygen and a range of other nonmetal elements, connecting metallic mixing with diverse compound chemistries. Figure~\ref{fig:2}c reveals distinct cell-size patterns: metallic configurations are concentrated mainly in the 96--191-atom range, while oxygen-containing configurations contribute strongly to the larger 192--319-atom cells. The bubble colours vary within and across these size ranges, showing that cell size and principal-pool mixing index characterize different aspects of the configurations. Figure~\ref{fig:2} thus establishes a benchmark spanning chemical families, balanced multielement compositions and varied cell sizes. Figure~\ref{fig:3} places these features in the context of the training references.

Figure~\ref{fig:3} compares the benchmark with MPtrj, sAlex and OMat24 at three levels: individual elements, their coexistence within a cell and cell size. In Fig.~\ref{fig:3}a, the four datasets share broad elemental coverage, but assign different relative frequencies to those elements. The benchmark places greater emphasis on several transition metals, including Ti, V, Cr, Mn, Zr, Nb and Mo, as well as halogen-containing compositions. Each element's bars are normalized by its largest occurrence fraction among the datasets, so their heights compare the relative prevalence of that element across datasets. This extends the occurrence map in Fig.~\ref{fig:2}b by showing how the benchmark's chemical emphasis differs from the references.

The strongest global contrast appears in Figs.~\ref{fig:3}b,c. The reference element-count distributions are dominated by cells with three or four species and fall rapidly towards higher counts. The benchmark instead occupies the multielement region, with its largest grouped fraction in the 10+ category. Figure~\ref{fig:3} counts all species in a cell, while Fig.~\ref{fig:2}a characterizes the designated principal mixing pool. In parallel, the benchmark cell-size distribution extends broadly through the larger-cell region. MPtrj retains a longer large-cell tail than sAlex and OMat24, but its probability mass still declines strongly with increasing cell size. Thus, shared elemental coverage coexists with a substantial shift in how many elements and atoms occur together. Read together, Figs.~\ref{fig:2} and~\ref{fig:3} define the transfer task: evaluating pretrained force fields on the multicomponent combinations and cell sizes emphasized by compositional design. The following energy, force, stress and derived-property comparisons quantify their performance in this setting, before the local-environment analyses resolve the associated chemical and geometric error patterns.

\subsection{Unified evaluation of force-field models}

\begin{figure}[!t]
\centering
\includegraphics[width=\textwidth]{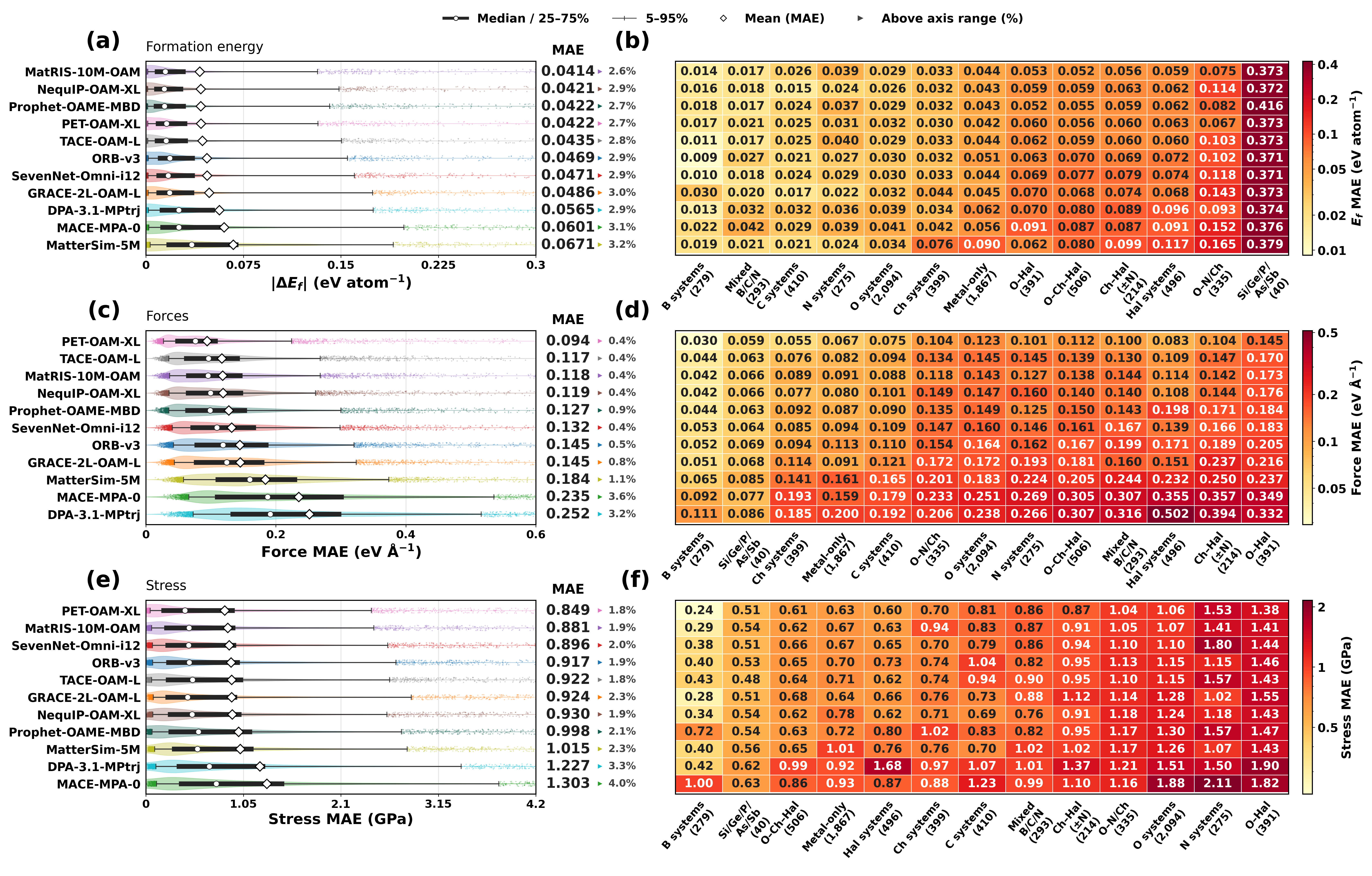}
\caption{\textbf{Energy, force and stress errors for eleven pretrained force fields.} a,c,e, Structure-level distributions of formation-energy error, Cartesian force MAE and six-component stress MAE, respectively, for the common set of 7,599 configurations. Symbols and intervals identify the mean, median, interquartile interval and 5th--95th percentile interval as indicated in the figure. Values beyond the visible axes remain in the numerical summaries. b,d,f, Arithmetic mean structure-level errors within each disjoint chemical class, using the same model predictions and reference labels. Model rows run from lower to higher full-cohort mean error. Chemical-class columns run from lower to higher class mean error averaged across models, with a separate order for each property. Class counts appear below the chemical-class names. Ch denotes S, Se or Te. Hal denotes F, Cl, Br or I. The class-specific samples partition the full cohort. Formation energies use aligned model and DFT unary references, whereas force and stress errors are computed directly from the DFT reference labels.}\label{fig:4}
\end{figure}

Formation energies use model-aligned unary references, whereas forces and stresses are compared directly against DFT reference labels. Across the eleven models, mean formation-energy errors range from approximately 0.041 to 0.067~eV atom$^{-1}$ (Fig.~\ref{fig:4}), with MatRIS-10M-OAM giving the lowest mean at 0.0414~eV atom$^{-1}$. The leading formation-energy models are closely grouped. Mean force MAEs span a wider relative range, from 0.0941~eV~\AA$^{-1}$ for PET-OAM-XL to 0.2516~eV~\AA$^{-1}$ for DPA-3.1-MPtrj. Mean stress MAEs range from 0.849 to 1.303 GPa, with PET-OAM-XL again giving the lowest mean. The structure-level distributions in Fig.~\ref{fig:4}a,c,e show how the medians and upper tails differ from these pooled means. For example, PET-OAM-XL has a force-error median of 0.0763~eV~\AA$^{-1}$ and a 95th percentile of 0.2241~eV~\AA$^{-1}$, compared with 0.1913 and 0.5158~eV~\AA$^{-1}$ for DPA-3.1-MPtrj.

The chemical-class heatmaps in Fig.~\ref{fig:4}b,d,f broadly rise from the upper left to the lower right because model rows and class columns are each ordered by mean error. The horizontal progression is shared across models, indicating recurring chemical-class contrasts alongside differences in overall model accuracy. In the force comparison, the higher-error region includes O--Hal, Ch--Hal and Hal systems, while B and Ch systems have lower class means. These compositional contrasts must be distinguished from the general sensitivity to short contacts. O--Hal systems combine oxygen and halogen species, unlike the Hal and Ch classes. Their median force error remains higher than that of Hal systems in all eleven models within each of three relative-contact intervals spanning 0.75--0.90. The contrast therefore persists among structures with similar values of this geometric descriptor. Conversely, some of the pooled Ch--Hal versus Ch contrast decreases within these intervals. The result directs attention to specific combinations of chemical species and their coordination environments, rather than anion mixing or short contacts alone, motivating the local chemical and elemental-response analyses below.

\begin{figure}[!t]
\centering
\includegraphics[width=\textwidth]{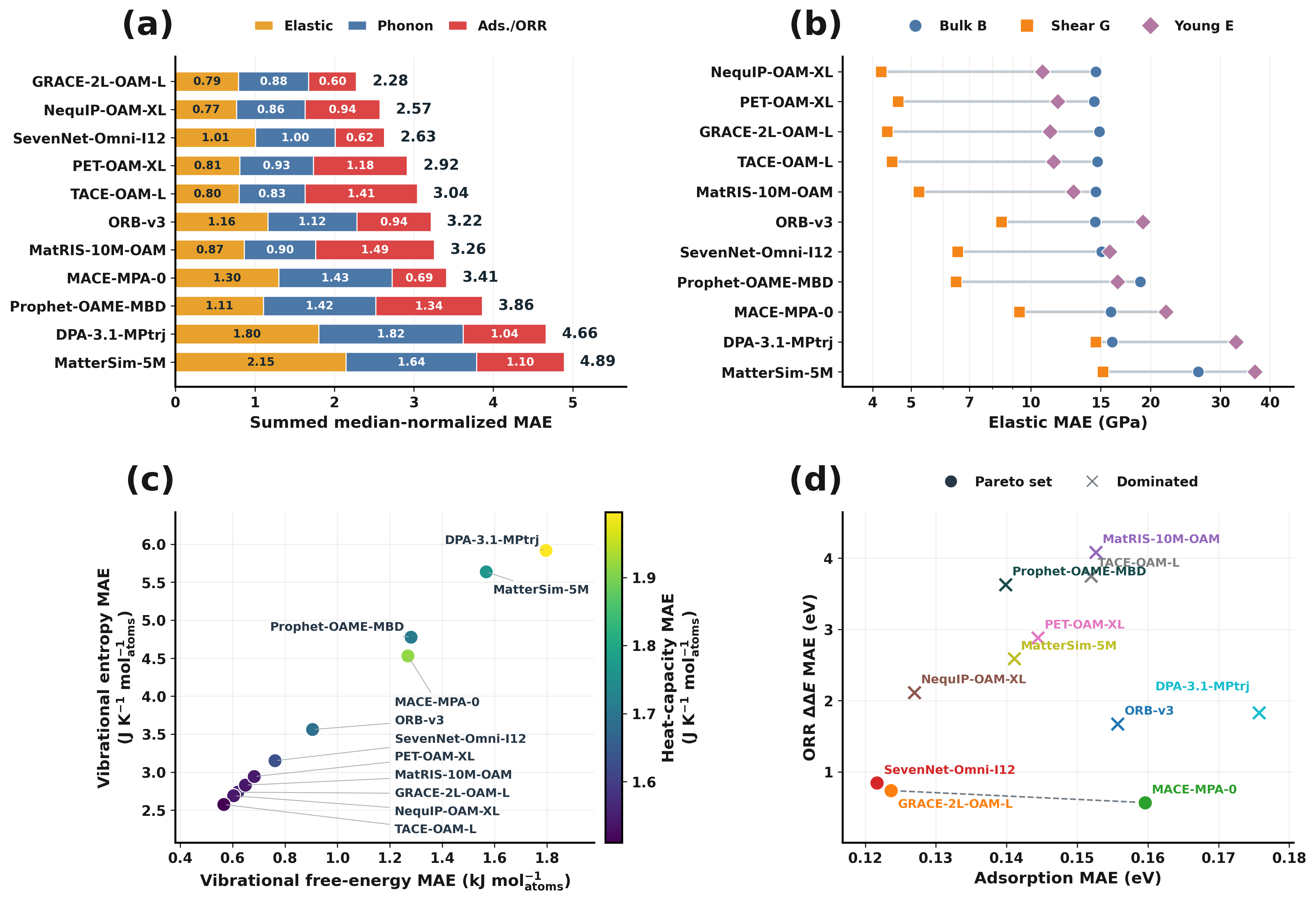}
\caption{\textbf{Comparison across elastic, vibrational and adsorption-related properties.} a, Sum of three equally weighted family-level normalized error scores for eleven models, ordered from lowest to highest total. Each property MAE is divided by its median across the eleven models, and properties are averaged within each family before the three family scores are summed. Orange, blue and red show the elastic, vibrational and surface-property contributions. Labels inside the bars give family scores and labels at their ends give totals. b, Bulk-, shear- and Young's-modulus MAEs on the common 480-structure positive-modulus subset. Nine structures with nonpositive DFT moduli and two additional structures with nonpositive model moduli are excluded from the elastic comparison for every model. Circles, squares and diamonds denote bulk, shear and Young's moduli. Grey lines connect the three errors for each model, and model rows are ordered by their mean rank across the three moduli. The horizontal axis is logarithmic. c, Vibrational free-energy MAE (kJ mol-atoms$^{-1}$) versus entropy MAE (J K$^{-1}$ mol-atoms$^{-1}$) at 300 K, with colour indicating heat-capacity MAE (J K$^{-1}$ mol-atoms$^{-1}$). All eleven models use the same 394 structures with converged DFT displacement calculations and protocol-matched model results. Values are normalized by atom count, and DFT and model spectra use the same positive-frequency integration rule. d, Centred O/OH adsorption-energy-proxy MAE for 8,168 entries versus pure-site-referenced OH reaction-energy-difference MAE for 726 entries. SevenNet-Omni-I12 minimizes the adsorption-proxy error, MACE-MPA-0 minimizes the OH energy-difference error, and GRACE-2L-OAM-L provides a third nondominated combination. Filled circles identify these three models. Crosses identify the other eight models, each outperformed on both errors by at least one nondominated model. The dashed line connects the nondominated points.}\label{fig:5}
\end{figure}

Figure~\ref{fig:5} extends the static comparison in Fig.~\ref{fig:4} to elastic responses, vibrational summaries and surface-energy differences, using 480 elastic structures and 394 vibrational structures. With equal weights for the three property families, GRACE-2L-OAM-L has the lowest summed normalized error (2.28), followed by NequIP-OAM-XL and SevenNet-Omni-I12 (Fig.~\ref{fig:5}a). The family scores reveal distinct strengths: NequIP-OAM-XL leads the elastic comparison, TACE-OAM-L leads the vibrational comparison and GRACE-2L-OAM-L leads the combined surface-property comparison. PET-OAM-XL, which has the lowest mean force and stress errors in Fig.~\ref{fig:4}, ranks 4 in this aggregate comparison. Its elastic and vibrational scores remain relatively low, while its larger surface-property score raises the total. Static accuracy therefore provides incomplete guidance for selecting a model for a specific derived property.

The elastic comparison shows greater relative separation between models for shear and Young's moduli than for the bulk modulus (Fig.~\ref{fig:5}b). The largest-to-smallest MAE ratios are 3.6 and 3.4, respectively, compared with 1.8 for the bulk modulus. NequIP-OAM-XL gives the lowest shear- and Young's-modulus MAEs. DPA-3.1-MPtrj illustrates the distinction: its bulk-modulus MAE is 16.0 GPa, within the range of most models, while its shear- and Young's-modulus MAEs reach 14.6 and 32.8 GPa. Young's modulus is derived from the bulk and shear moduli, so its increase here accompanies the poorer shear response. In the vibrational comparison, free-energy and entropy errors rise together from the lower-left cluster towards DPA-3.1-MPtrj and MatterSim-5M in the upper right (Fig.~\ref{fig:5}c). TACE-OAM-L gives the lowest error for all three vibrational quantities. Heat-capacity errors span only 1.51--2.00~J K$^{-1}$ mol-atoms$^{-1}$, a narrower relative range than the free-energy and entropy errors. Elastic constants and phonon force constants probe energy curvature under strain and atomic displacement, respectively, adding response information beyond the static energies, forces and stresses in Fig.~\ref{fig:4}.

The surface-property plane shows a much larger relative spread along the OH reaction-energy-difference axis than along the adsorption-proxy axis (Fig.~\ref{fig:5}d). SevenNet-Omni-I12 has the lowest adsorption-proxy MAE (0.122 eV), whereas MACE-MPA-0 has the lowest OH energy-difference MAE (0.569 eV). GRACE-2L-OAM-L combines an adsorption-proxy MAE close to SevenNet's (0.124 eV) with a smaller OH energy-difference error (0.737 eV). These three models define the nondominated set. The other eight models can each be outperformed on both errors by at least one model in this set, so the plane does not imply an unavoidable trade-off for all models. The comparison with Fig.~\ref{fig:4} is especially clear for MatRIS-10M-OAM: despite its lowest mean formation-energy error in the static benchmark, it has the largest OH energy-difference error here (4.08 eV). PET-OAM-XL's leading static force and stress accuracy likewise does not yield the lowest surface-property errors, while MACE-MPA-0 leads the OH comparison despite its larger elastic and vibrational errors. These changes in ordering complement the chemical-system contrasts in Fig.~\ref{fig:4}. Bulk static labels, deformation and vibrational responses, and surface-energy differences test different aspects of the predicted energy landscape, making both the intended chemical environment and the target observable necessary parts of model evaluation.

\subsection{Local-environment coverage and qualitative error variation}

\begin{figure}[!t]
\centering
\includegraphics[width=\textwidth]{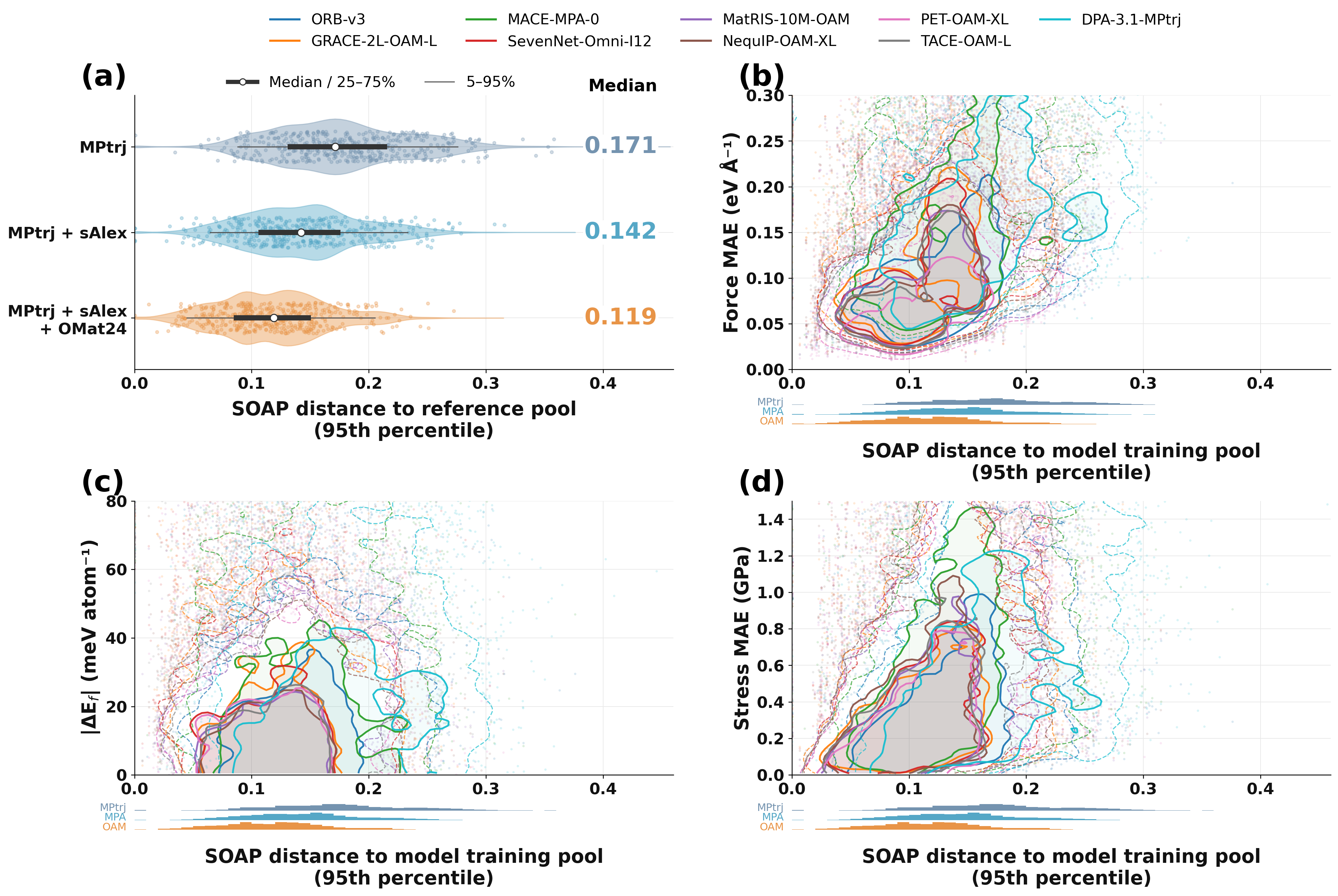}
\caption{\textbf{Training-reference distance and prediction error across 7,599 structures.} a, Distributions of the structure-level 95th-percentile atomic SOAP distance to three nested reference pools. Points show a subset of structures for clarity. Distributions and quantiles use all structures. White markers denote medians, thick lines the interquartile ranges and thin lines the 5th--95th percentiles. b--d, Joint distributions of model-assigned reference distance with force MAE, formation-energy error and stress MAE, respectively, for nine models. Solid and dashed contours enclose the 50\% and 80\% highest-density mass. Faint points show lower-density observations. Histogram strips show reference-pool distance frequencies. Reference assignments use MPtrj for DPA-3.1-MPtrj, MPtrj plus sAlex for ORB-v3 and MACE-MPA-0, and all three reference datasets for the other plotted models. MatterSim-5M and Prophet-OAME-MBD are absent from this comparison.}\label{fig:6}
\end{figure}

To examine the environmental dependence of the static prediction errors, we calculate for each atomic environment the distance to the nearest reference environment with the same central element and summarize each structure by the 95th percentile of its atomic distances. The median structure-level distance decreases from 0.171 for MPtrj to 0.142 for MPtrj plus sAlex, and to 0.119 after adding OMat24 (Fig.~\ref{fig:6}a). The decrease reflects the closer reference environments supplied by the expanded pools. The model-resolved comparison uses nine models evaluated on the same 7,599 structures, with reference pools assigned to approximate their training coverage. ORB-v3 and MACE-MPA-0 use MPtrj plus sAlex, DPA-3.1-MPtrj uses MPtrj, and the remaining plotted models use all three reference datasets.

As reference distance increases, the force and stress distributions extend towards larger errors (Fig.~\ref{fig:6}b,d). This trend is present within every model, rather than arising only from differences between the reference pools. Comparing each model's highest- and lowest-distance quartiles, the median force error increases by a factor of 1.59--2.49 and the median stress error by 2.32--3.94. The formation-energy plane shows a much weaker directional change (Fig.~\ref{fig:6}c). Large and small energy errors coexist over much of the populated distance range, and the within-model Spearman correlations span only $-0.12$ to 0.11, compared with 0.34--0.50 for force and 0.31--0.42 for stress. Thus, increasing reference distance is more consistently associated with errors in force and stress than with formation-energy error.

This property dependence is consistent with energy and its derivatives testing different aspects of the predicted energy landscape. Agreement in energy at a configuration does not ensure accurate gradients under atomic displacement or cell strain. Reference distance therefore indicates a greater risk of force and stress errors without defining a unique error level. The broad vertical spread at similar distances, together with differences between models assigned the same reference pool, shows that coverage alone does not determine accuracy. A scalar nearest-reference distance summarizes the magnitude of an environmental difference but does not identify which chemical or geometric changes are responsible for the error. These observations connect the model-dependent static accuracy in Fig.~\ref{fig:4} to the subsequent analyses of local chemistry, geometric heterogeneity and distance directionality.

\subsection{Fine-grained structural and elemental error analysis}

To resolve the chemical and geometric variation behind the broad error spread in Fig.~\ref{fig:6}, we train an atom-based regressor to predict the mean atomic Cartesian force error across the eleven evaluated force fields. Held-out atomic predictions are averaged within each structure.

\begin{figure}[!t]
\centering
\includegraphics[width=\textwidth]{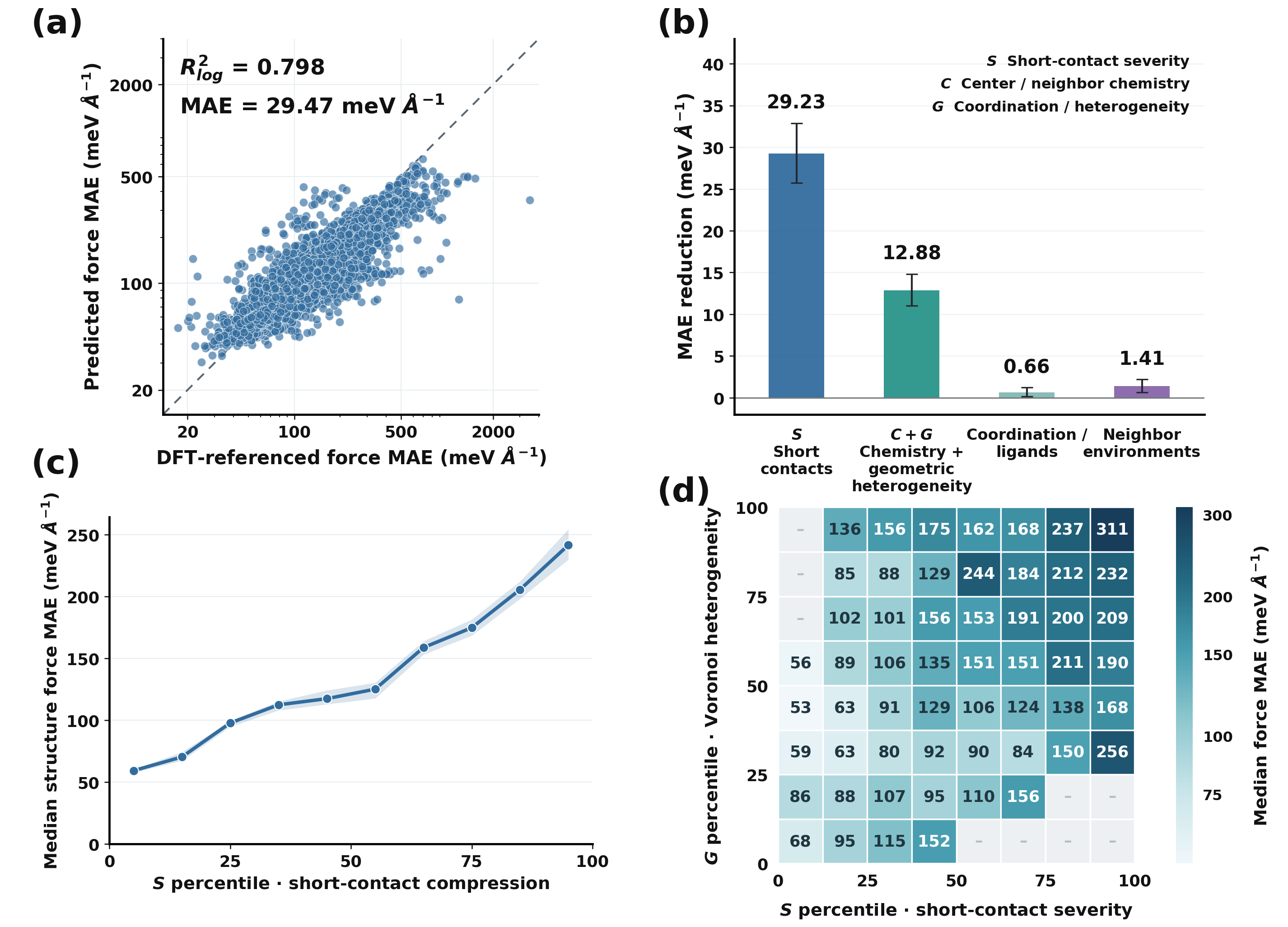}
\caption{\textbf{Structure-based force-error prediction and chemical--geometric contributions.} a, Out-of-fold predicted versus DFT-referenced force MAE for the atom-based error regressor, averaged to the structure level across 7,599 configurations. The target is the mean atomic error across eleven force fields. Prediction MAE is 29.47~meV~\AA$^{-1}$ and $R^{2}_{\log} = 0.798$. b, Paired reductions in held-out prediction MAE: constant to short-contact-only model, then short-contact-only to full model. The smaller bars compare coordination/ligand and neighbour-environment extensions with the base atom model. Error bars show 95\% intervals from 1,000 paired group bootstrap resamples. c, Median structure force MAE across all valid benchmark structures, grouped into ten percentile bins of short-contact compression. Points show bin medians and shaded bands show 95\% bootstrap intervals. d, Median structure force MAE in percentile bins of short-contact severity and mean Voronoi heterogeneity for 7,583 valid structures. Grey cells contain fewer than 20 structures. S denotes short-contact severity, C central/neighbour chemistry, and G coordination and geometric heterogeneity.}\label{fig:7}
\end{figure}

The complete descriptor model achieves a structure-level prediction MAE of 29.47~meV~\AA$^{-1}$ and a logarithmic coefficient of determination of 0.798 (Fig.~\ref{fig:7}a). This MAE measures how accurately the regressor estimates force-error magnitudes. It does not measure an improvement to the underlying force fields. On the same held-out groups, the constant baseline and short-contact-only predictor have MAEs of 71.57 and 42.35~meV~\AA$^{-1}$, respectively. Including the full chemical and geometric descriptors lowers the latter value to the 29.47~meV~\AA$^{-1}$ shown in Fig.~\ref{fig:7}a. Figure~\ref{fig:7}b displays the corresponding reductions of 29.23 and 12.88~meV~\AA$^{-1}$. The second reduction is 30.4\% of the short-contact-only prediction MAE, with a grouped-bootstrap 95\% confidence interval of 11.04--14.81~meV~\AA$^{-1}$ for the absolute reduction. Thus, short contacts capture substantial predictive information, while central-element identity, neighbouring species and local geometry provide additional information.

Figures~\ref{fig:7}c,d examine the measured force errors directly to show which environments carry higher errors. In Fig.~\ref{fig:7}c, structures are ordered by short-contact severity, defined from near-neighbour distances relative to the sums of covalent radii. A higher percentile indicates more pronounced short contacts within this benchmark, rather than a specified percentage of mechanical compression. Across ten percentile bins, the median structure force error rises from approximately 60 to 240~meV~\AA$^{-1}$. The increase is gradual through the middle range and becomes more pronounced towards the most severe short contacts. Because each point represents a bin median, the trend reflects a shift in typical error across the population rather than a few extreme structures. It identifies short-contact severity as a useful geometric indicator of elevated force-error risk.

Figure~\ref{fig:7}d adds a second geometric coordinate, the structure-averaged Voronoi heterogeneity, which measures the spread of local neighbour distances weighted by Voronoi face areas. The overall error landscape rises from the lower left, where short contacts are less severe and neighbour distances are more uniform, towards the upper right, where pronounced short contacts coexist with more heterogeneous local environments. This two-dimensional view places the trend in Fig.~\ref{fig:7}c within the broader variation of coordination geometry. At comparable short-contact percentiles, many columns also show higher median errors towards greater heterogeneity, although the increase is not monotonic in every column. The upper-right region therefore identifies environments combining short contacts and uneven neighbour-distance distributions as a recurring concentration of high errors. Together with the descriptor comparison in Fig.~\ref{fig:7}b, this pattern motivates resolving the identities of the central and neighbouring elements and the geometry of their coordination shells in the following analyses.

\begin{figure}[!t]
\centering
\includegraphics[width=\textwidth]{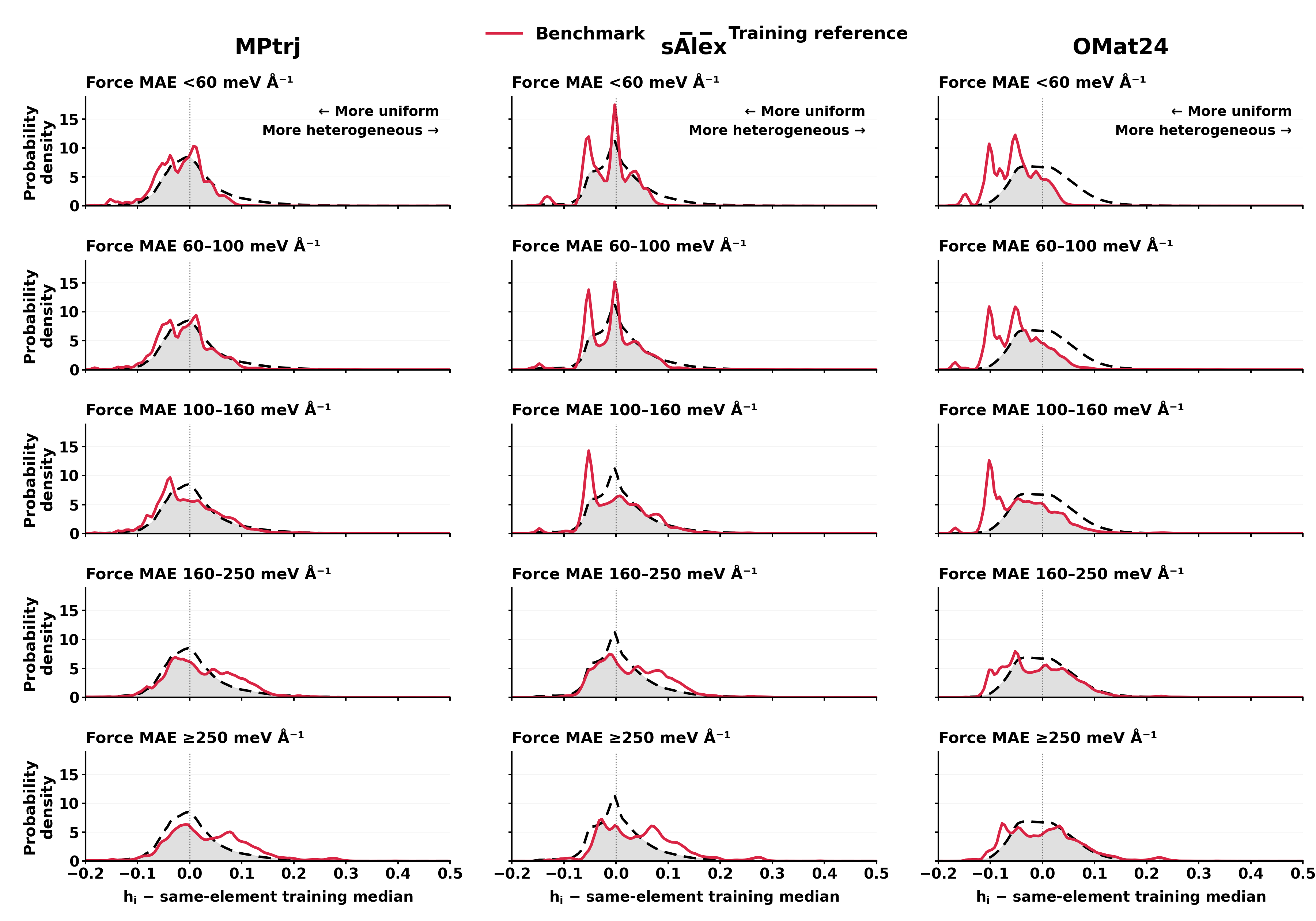}
\caption{\textbf{Element-aligned Voronoi heterogeneity across force-error groups.} Columns compare the benchmark with MPtrj, sAlex and OMat24, and rows show five groups of structure-level mean force MAE. The horizontal coordinate is the site heterogeneity $h_{i}$ minus the training-reference median for the same central element. Here $h_{i}$ is the Voronoi-face-area-weighted mean absolute deviation of neighbour distances divided by the weighted mean distance. Red curves denote the benchmark and dashed black curves the training reference. Grey shading marks their overlap. Curves give equal weight to structures within a central element and then to each of the 56 commonly supported elements. Of 7,599 attempted benchmark tessellations, 7,583 are valid. Grouping uses the ten-model mean error, excluding Prophet-OAME-MBD. The distributions are dimensionless and are smoothed only for display.}\label{fig:8}
\end{figure}

Figure~\ref{fig:7} associates larger force errors with more heterogeneous local geometry. Figure~\ref{fig:8} asks how these environments compare with those represented in the training references. The site descriptor $h_{i}$ measures the Voronoi-face-area-weighted spread of neighbour distances relative to their weighted mean. Larger values indicate more uneven distances within a coordination environment. For each central element, we subtract its median $h_{i}$ in the chosen reference dataset. Zero therefore denotes that element's reference median, with negative and positive values indicating more uniform and more heterogeneous environments, respectively. The three columns compare the same benchmark groups with MPtrj, sAlex and OMat24. The five rows progress from low to high structure-level force error, using the mean across ten force fields. The comparison includes 7,583 valid structures and 56 commonly supported central elements, with equal structure weights within each element and equal element weights.

Read vertically, all three columns show a redistribution towards greater heterogeneity as force error increases. Low-error groups concentrate more strongly in relatively uniform environments, whereas high-error groups develop broader positive-side shoulders and tails. The persistence of this pattern across the three references connects the structure-level trend in Fig.~\ref{fig:7}d to the underlying distributions of local coordination environments. Read horizontally, however, the interpretation relative to training coverage changes. Against MPtrj and sAlex, the highest-error groups have excess probability density on the more heterogeneous side of the reference distributions. Against OMat24, the low-error benchmark groups lie mainly to the left of the reference centre, while the high-error groups overlap it more closely. The benchmark environments are unchanged between columns. Their shifted positions reflect the different same-element reference medians.

\begin{figure}[!t]
\centering
\includegraphics[width=\textwidth]{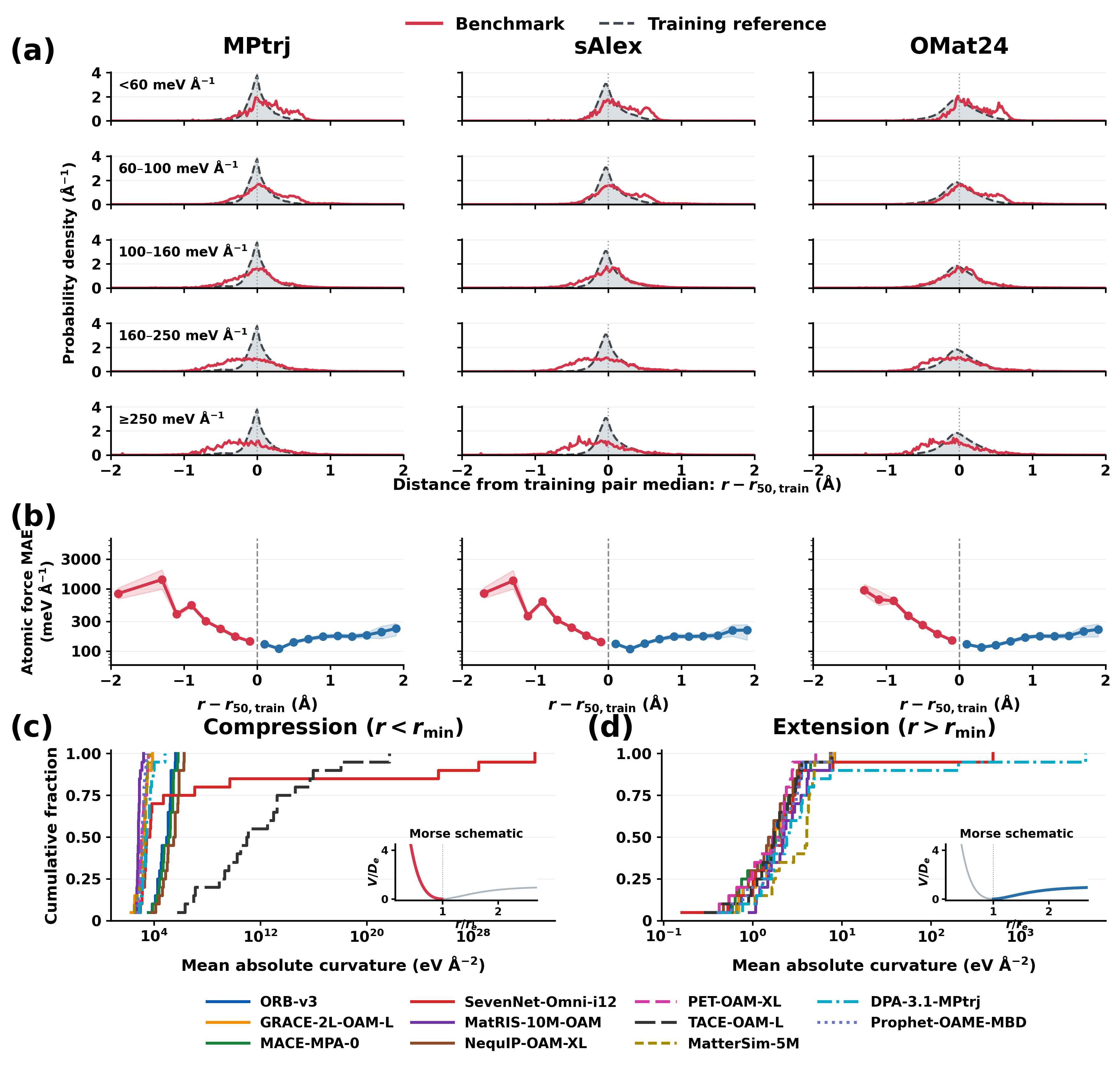}
\caption{\textbf{Asymmetric force errors and interaction responses under compression and extension.} a, Distributions of interatomic-distance deviations from the training-reference pair median in five structure-level force-MAE groups. Columns use MPtrj, sAlex and OMat24. Red curves represent the benchmark and grey dashed curves the training references. Each benchmark structure is normalized within its element pair before averaging, with equal pair weights. The 125 common pairs contribute contacts from 6,405 of 7,599 structures. Full-cohort group counts are 912, 1,874, 2,154, 1,779 and 880, with corresponding common-pair counts of 695, 1,648, 1,734, 1,581 and 747. b, Atomic force error, averaged across eleven models, versus signed distance deviation. Red and blue indicate compression and extension relative to the reference median. Shading shows 95\% structure-bootstrap intervals from 250 resamples. c,d, Empirical cumulative distributions of mean absolute curvature in the compressed (c) and extended (d) regions of the predicted pairwise interactions, evaluated using two-atom configurations of each element. Colours and line patterns identify the eleven models in the shared legend. Each distribution contains the same 20 elements with complete scans and interior energy minima in the 1--3~\AA{} search interval. The ordinate gives the fraction of elements at or below each curvature value. Curvature is evaluated from radial-force differences over 0.10--5.00~\AA{}. The logarithmic horizontal axes have different ranges. Insets show the corresponding branches of a dimensionless Morse potential as schematics. Panels a,b use the training-reference pair median as the distance origin, while c,d use each model's predicted energy minimum to separate the two branches.}\label{fig:9}
\end{figure}

\begin{figure}[!t]
\centering
\includegraphics[width=\textwidth]{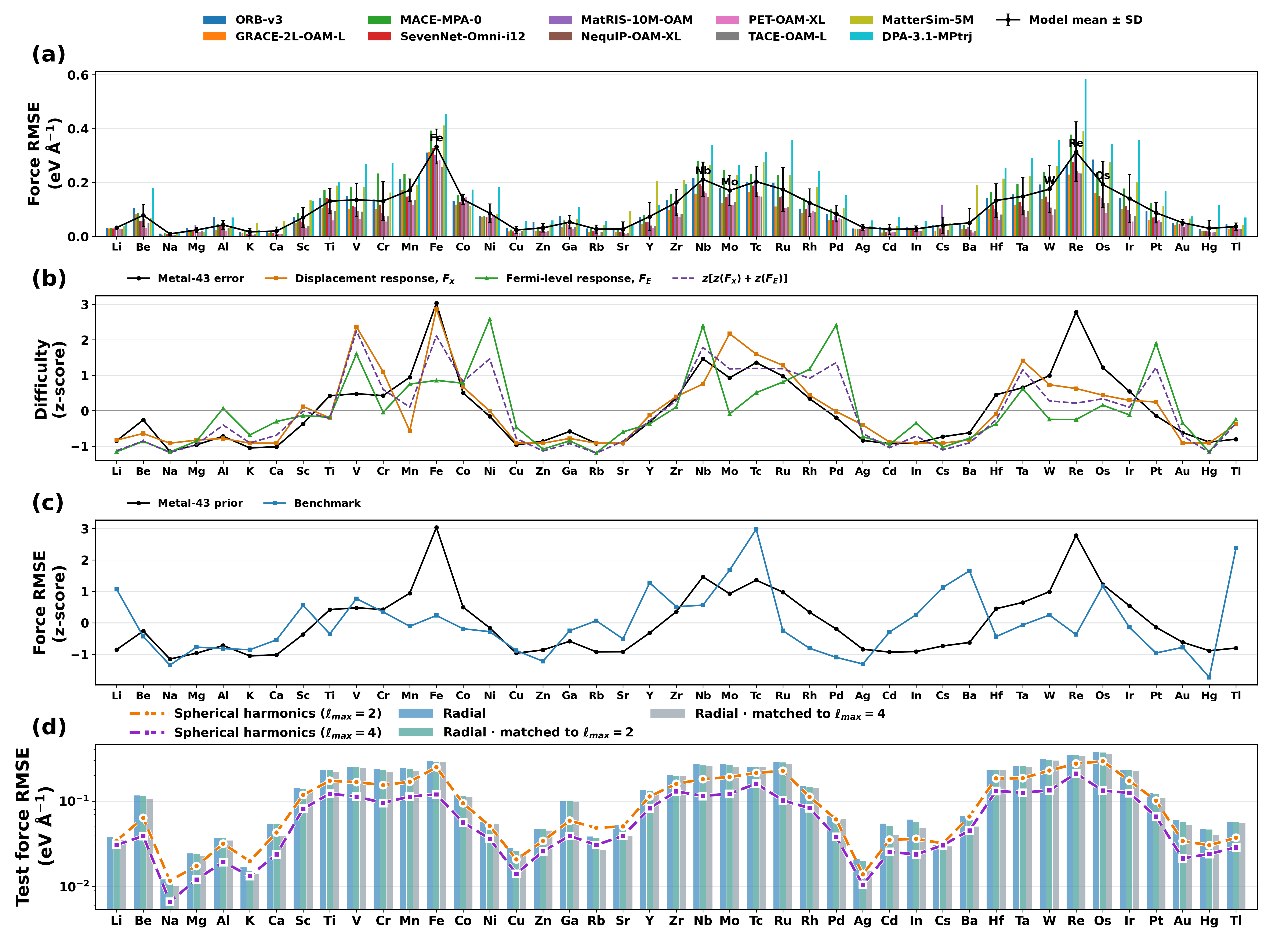}
\caption{\textbf{Elemental fitting difficulty and its transfer to multicomponent environments.} a, Model-specific force RMSEs for 43 elemental metals in the independent Metal-43 reference. The black line shows the mean $\pm$ standard deviation across ten models. b, Standardized reference error, displacement response $F_{x}$, Fermi-level response $F_{E}$ and the restandardized sum $z[z(F_{x}) + z(F_{E})]$. $F_{x}$ measures the sensitivity of the occupied electronic eigenvalue sum to atomic displacement. $F_{E}$ measures its variation with Fermi-level shifts. The recurring high-error regions in a co-vary with these complementary response measures. c, Comparison of the Metal-43 profile with benchmark element-specific RMSEs. Benchmark RMSEs pool squared Cartesian errors by element for each model and are then averaged across models. Each series is standardized separately. Local chemistry and geometry can change the detailed ranking and error amplitude. d, Element-specific test force RMSEs for radial and spherical-harmonic models trained on a sampled Metal-43 subset, averaged over two random seeds. Blue bars denote the radial baseline. Teal and grey bars denote radial models with parameter counts matched to the second- and fourth-order spherical-harmonic models, respectively. Orange and purple dashed lines denote spherical-harmonic models with maximum degrees $\ell_{\max} = 2$ and 4, respectively. All models use identical training, validation and test splits.}\label{fig:10}
\end{figure}

This distinction is quantified by the distribution overlap. For OMat24, overlap increases from 0.57 in the lowest-error group to 0.84 in the highest-error group. The corresponding highest-error overlaps are 0.76 for MPtrj and 0.72 for sAlex. Thus, high force errors can coexist with substantial reference overlap in this geometric descriptor. Greater local heterogeneity and greater departure from a reference distribution are distinct features of the error landscape. The former recurs across the error groups, while the latter depends on the reference dataset. These one-dimensional overlaps do not establish equivalence of the full chemical environments or rank the accuracy of models trained on each reference. They instead identify what the heterogeneity descriptor leaves unresolved: it measures the spread of neighbour distances but does not distinguish whether particular element pairs are compressed or extended. Figure~\ref{fig:9} examines this directionality before Fig.~\ref{fig:10} addresses element-dependent response.

Figure~\ref{fig:8} measures how uneven local neighbour distances are, but does not distinguish shorter from longer contacts. Figure~\ref{fig:9} resolves this directionality by referencing each interatomic distance to the training median for the same element pair. Negative and positive deviations denote relative compression and extension, rather than imposed mechanical strains. Figure~\ref{fig:9}a compares five force-error groups using 125 pairs shared across all groups and three references, with contributions from 6,405 structures. Read from low to high error, the benchmark distributions broaden and acquire more weight on the negative side. Low-error groups can retain a substantial extension-side shoulder, whereas the highest-error groups show a pronounced compression-side excess. This contrast recurs with MPtrj, sAlex and OMat24, despite differences in their reference distributions. Thus, the distributions suggest that the direction of a distance deviation matters in addition to its magnitude.

Figure~\ref{fig:9}b tests this interpretation by relating atomic force error directly to signed distance deviation. Errors are lowest near the reference median. Moving towards shorter separations produces a steep increase, whereas moving towards longer separations produces a much gentler rise. The similar shapes across the three references show that the contrast is not specific to one choice of reference median. After matching element pairs and absolute distance-offset bins, compression-side mean errors are 234--249~meV~\AA$^{-1}$, compared with 128--129~meV~\AA$^{-1}$ on the extension side, giving ratios of 1.81--1.95. Figure~\ref{fig:9}a therefore identifies which contacts become more prevalent in high-error structures, while Fig.~\ref{fig:9}b establishes their unequal error levels at comparable distance deviations. Together, they refine the short-contact trend in Fig.~\ref{fig:7} and the unsigned heterogeneity comparison in Fig.~\ref{fig:8}: departures towards shorter separations carry substantially greater errors than comparable departures towards longer separations.

To examine the interaction responses associated with this asymmetry, we follow UniFFBench\cite{mannan2026} and scan model-predicted two-atom configurations for the same 20 supported elements. Figures~\ref{fig:9}c,d summarize the mean absolute curvature on either side of each model's predicted energy minimum. Curvature measures how rapidly the predicted radial force changes with distance. At a fixed cumulative fraction, a curve further right indicates larger curvature. All eleven models have a higher median curvature on the compression branch. The compression distributions also separate much more strongly between models, whereas the extension distributions cluster more closely over most elements. TACE-OAM-L is markedly shifted towards large compression curvatures, and SevenNet-Omni-I12 has a particularly long upper tail. This contrast is consistent with a steep repulsive branch and a gentler extension branch, as illustrated schematically by the Morse potential.

Figure~\ref{fig:9} thus reveals a consistent directional contrast: compression is associated with larger force errors in the benchmark and with steeper predicted force responses in the dimer scans. The compressed branch also shows greater variation between models. These observations connect the short-contact trend in Fig.~\ref{fig:7} and the heterogeneous distance distributions in Fig.~\ref{fig:8} with the directional character of the interaction response. The next question is how this response varies with elemental identity, motivating the electronic-response and elemental fitting analysis in Fig.~\ref{fig:10}.

Figure~\ref{fig:9} identifies the direction of distance changes associated with larger force errors. Figure~\ref{fig:10} adds the elemental dimension by asking which elements are consistently difficult to fit and how that difficulty relates to electronic response. In the independent Metal-43 reference, the ten evaluated models share recurring high-error regions around Fe, Nb, Mo/Tc, W, Re and Os (Fig.~\ref{fig:10}a). Although their absolute errors differ, the model mean preserves these peaks, indicating an elemental pattern shared across architectures. Figure~\ref{fig:10}b compares this pattern with two complementary measures\cite{geng2026}. The displacement response $F_{x}$ measures how strongly the sum of occupied electronic eigenvalues changes when atoms move. The Fermi-level response $F_{E}$ measures how that sum changes when the boundary of occupied states shifts. After standardization across elements, these quantities can be compared with the relative force-error profile. Fe has a pronounced displacement-response peak, while Nb has a strong Fermi-level response. The differing profiles show why the two measures provide complementary information. Their standardized sum summarizes the joint pattern, connecting elemental fitting difficulty with both atomic-motion sensitivity and electronic response near the Fermi level.

Figure~\ref{fig:10}c asks whether this elemental pattern remains visible when the same elements occur in multicomponent environments. The Metal-43 and benchmark error profiles are standardized separately, so the comparison concerns relative difficulty across elements. Several low-error regions and the elevated Mo/Tc region recur in both profiles. At the same time, the prominent Fe and Re peaks in Metal-43 become less pronounced in the multicomponent benchmark. Thus, the elemental pattern carries over in part and is reshaped by the surrounding chemistry and coordination. This connects the electronic-response analysis with Figs.~\ref{fig:7}--\ref{fig:9}: the error associated with a local environment reflects both its geometry and the identities of the interacting elements.

Figure~\ref{fig:10}d then tests whether resolving angular information improves the fitting of these elemental systems. Radial models and spherical-harmonic models with maximum degrees $\ell_{\max} = 2$ and 4 use the same sampled Metal-43 data and identical training, validation and test splits. Wider radial networks provide a parameter-matched control for each angular model. Relative to its matched radial control, the $\ell_{\max} = 2$ model lowers test force RMSE for 38 of 43 elements, with a mean element-wise reduction of 19.1\%. The $\ell_{\max} = 4$ model improves 40 of 43 elements, with a mean reduction of 39.8\%. Comparing the two angular models directly, $\ell_{\max} = 4$ improves all 43 elements and reduces RMSE by 32.1\% on average. For the high-error elements Fe, Nb, Mo, Tc, W, Re and Os, its mean reduction is 38.4\% relative to $\ell_{\max} = 2$ and 50.8\% relative to its parameter-matched radial control. These percentages are calculated for each element before averaging. The comparison shows that explicit angular information improves elemental force fitting at matched parameter counts, with particularly large gains for several of the difficult transition metals identified in Fig.~\ref{fig:10}a.

\section{Discussion}\label{sec:discussion}

The benchmark tests the transfer of pretrained force fields to multicomponent configurations motivated by composition-based materials design. The eleven models show substantial differences in accuracy on the same configurations, with distinct performance patterns across static predictions and elastic, vibrational and adsorption-related properties. The error analysis proceeds from environmental coverage to local geometric heterogeneity, then resolves the roles of interatomic-distance directionality and elemental response. Distances to training-reference environments describe coverage differences and their qualitative association with error, while substantial error variation remains at similar distances. Element-aligned Voronoi distributions show that higher-error configurations extend towards greater heterogeneity within their coordination shells. OMat24 provides broader coverage of these heterogeneous environments than MPtrj and sAlex, including substantial overlap with the highest-error configurations. This overlap motivates a finer analysis of the distance variations and chemical environments underlying the heterogeneity distributions.

Separating interatomic-distance deviations by direction reveals a pronounced asymmetry within this broader geometric variation. Force errors remain low near the training-reference pair median, rise steeply on the compression side and increase more weakly on the extension side. After matching element pairs and absolute distance-offset bins, compression-side mean force errors are 1.81--1.95 times extension-side errors. The model-predicted pairwise interaction curves also show greater curvature on the compressed branch, providing a qualitative physical context for the directional contrast. Elemental response adds a chemical dimension to this geometric analysis. In Metal-43, fitting difficulty is associated with electronic band-energy responses to atomic displacement and Fermi-level shifts, and its overall trend persists in element-resolved multicomponent errors. Parameter-matched comparisons on Metal-43 further show that explicit angular representations improve elemental force fitting. The fourth-order spherical-harmonic model reduces element-wise force RMSE by 39.8\% on average relative to its radial control, with a larger mean reduction of 50.8\% for Fe, Nb, Mo, Tc, W, Re and Os. These results support angular representation as a route to improving elemental force fitting. Distance directionality and elemental response thus resolve specific error patterns within the broader coverage and heterogeneity comparisons. The error predictor also quantifies the information added by local chemistry and geometry beyond short-distance descriptors: prediction MAE falls from 42.35 to 29.47~meV~\AA$^{-1}$.

These findings point to a more specific challenge than increasing the volume of training data: identifying local environments that remain unreliable even when their training-reference distances suggest adequate coverage. Prospective data selection could combine chemical family, coordination-shell heterogeneity and the direction of element-pair distance deviations to identify high-risk configurations, while filtering out implausible structures rather than oversampling extreme contacts. Its benefit should be tested against random and coverage-based selection at matched DFT-label budgets, with entire chemical families held out to distinguish transfer from interpolation. Model development poses a separate question. Parameter-matched radial and angular architectures can test whether the elemental gains observed in Metal-43 carry over to multicomponent configurations and reduce compression-side errors. Finally, following errors through structure relaxation and property calculations would reveal whether improved force predictions translate into more reliable elastic, vibrational and adsorption-related properties. Together, these tests would define which combinations of data selection and representation improve transfer, and where pretrained force fields remain unreliable for compositional design.

\section{Methods}\label{sec:methods}

\subsection{Dataset construction and reference labels}

The benchmark contains 7,599 computational configurations generated through elemental substitution and anion mixing inspired by high-entropy compositional design. The elements participating in compositional mixing define the principal-element pool. Selected configurations contain at least five elements in this pool, each contributing 5--35\%, and satisfy the ideal mixing-index criterion

\begin{equation}
H_{\mathrm{pool}} = -\sum_{e} x_{e} \ln x_{e} \geq 1.5, \qquad \sum_{e} x_{e} = 1
\label{eq:1}
\end{equation}

\noindent where $x_{e} = n_{e}/N_{\mathrm{pool}}$, $n_{e}$ is the number of atoms of element $e$ assigned to the principal mixing pool in the structure-design record, and $N_{\mathrm{pool}} = \sum_{e} n_{e}$ is the total number of atoms in that pool. The sum runs over pool elements, $\ln$ denotes the natural logarithm, and $H_{\mathrm{pool}}$ is dimensionless. For a pool with $m$ equally represented elements, $H_{\mathrm{pool}} = \ln(m)$. The benchmark includes metal and intermetallic configurations, oxygen-containing compounds and oxygen-free compounds. Elemental occurrence is the fraction of structures containing an element. Comparisons with MPtrj, sAlex and OMat24 normalize this fraction by its maximum across datasets for each element. Element-count distributions include all species in each structure.

All eleven force fields are evaluated in their pretrained state on the same structures and DFT labels: MACE-MPA-0\cite{batatia2025}, ORB-v3\cite{rhodes2025}, GRACE-2L-OAM-L\cite{lysogorskiy2026}, SevenNet-Omni-I12\cite{kim2026}, NequIP-OAM-XL\cite{kavanagh2026}, PET-OAM-XL\cite{bigi2026}, TACE-OAM-L\cite{xu2026}, MatRIS-10M-OAM\cite{zhou2026}, MatterSim-5M\cite{yang2024}, DPA-3.1-MPtrj\cite{zhang2026} and Prophet-OAME-MBD\cite{kairosmaterials2026}. Formation-energy errors align model and DFT unary references using

\begin{equation}
\epsilon_{E} = \left| \frac{E_{M} - E_{\mathrm{DFT}}}{N} - \sum_{e} c_{e} \left( E_{M,e}^{\mathrm{unary}} - E_{\mathrm{DFT},e}^{\mathrm{unary}} \right) \right|
\label{eq:2}
\end{equation}

\noindent where $c_{e}$ is the atomic fraction of element $e$ in the full composition. The structure-level Cartesian force MAE is

\begin{equation}
\epsilon_{F,jm} = \frac{1}{3N_{j}} \sum_{i} \sum_{\alpha} \left| F_{m,ji\alpha} - F_{\mathrm{DFT},ji\alpha} \right|
\label{eq:3}
\end{equation}

\noindent where $N_{j}$ is the number of atoms and $\alpha$ indexes Cartesian components. Stress MAE averages the absolute differences over six independent components after unit and convention alignment. Overall MAEs weight structures equally. Distribution summaries use all structures, including values outside displayed axis limits. For the chemical-class force comparison, each atom's first shell contains neighbours within 1.25 times its nearest-neighbour distance. The short-contact ratio is the fifth percentile of these interatomic distances after dividing each distance by the sum of the corresponding covalent radii. Exploratory comparisons use intervals of width 0.05 over 0.75--0.95 and retain class comparisons with at least 20 structures in each interval. We compare class medians separately for each force field. Equal values of this descriptor do not imply identical physical bond strain or otherwise matched local environments.

The evaluated releases are the medium MACE-MPA-0 checkpoint, the ORB-v3 conservative infinite-neighbour MPA checkpoint dated 4 April 2025, GRACE-2L-OMAT-large-ft-AM, SevenNet-Omni-I12 with the MPA head, NequIP-OAM-XL, PET-OAM-XL v1.0.0, TACE-OAM-L, MatRIS-10M-OAM, MatterSim v1.0.0-5M, the DPA-3.1-MPtrj checkpoint dated 14 March 2025 and prophet-oame-mbd.pt. Prophet-OAME-MBD is the Matbench Discovery release\cite{kairosmaterials2026}.

\subsection{Derived-property evaluation}

Elastic properties are calculated with fixed ions using 25 strained and equilibrium configurations for each of 491 structures. Young's modulus is obtained from $E = 9BG/(3B + G)$. The comparison uses the common 480-structure subset with finite, positive bulk, shear and Young's moduli in DFT and all eleven models. Vibrational evaluation follows the matcalc workflow\cite{matcalccontributors} and uses Phonopy\cite{togo2023} for phonon post-processing. Vibrational properties are evaluated at 300~K on the common 394-structure subset with converged DFT calculations and protocol-matched results from all eleven models, drawn from 491 candidate structures. Raw phonon free energies in kJ per mole of cells and entropies and heat capacities in J K$^{-1}$ per mole of cells are divided by the number of atoms in the cell. The comparison therefore expresses free energies in kJ mol-atoms$^{-1}$, and entropies and heat capacities in J K$^{-1}$ mol-atoms$^{-1}$. DFT displacement calculations follow the pymatgen MPRelaxSet\cite{ong2013} electronic settings used by quacc\cite{rosen}, including the structure-dependent electronic convergence tolerance. The fixed-geometry phonon subset has no additional parent-force acceptance threshold. Force constants use paired positive and negative displacements of 0.015~\AA{} with matching supercell and phonon integration settings. Thermal sums integrate positive-frequency modes using a zero-frequency cutoff in both DFT and model spectra.

Adsorption-related evaluation uses 8,168 O/OH adsorption-proxy entries from the AgIrPdPtRu electrocatalyst dataset\cite{cao2025} and 726 OH reaction-energy-difference entries from the HEA ORR dataset\cite{huang2024}. These surface datasets are separately sourced from the static benchmark. For the adsorption proxy, DFT and model energies are separately centred within each adsorbate group before comparison. OH reaction-energy differences use the corresponding pure-site references. Each property MAE is divided by its median across models. Normalized errors are averaged within the elastic, vibrational and adsorption-related families and then summed with equal family weights. The adsorption comparison also identifies models that are not outperformed by another model on both plotted errors.

\subsection{Reference coverage and structural error prediction}

Local environments are represented by normalized SOAP descriptors\cite{bartok2013,himanen2020}, using a 4~\AA{} cutoff, $n_{\max} = 2$, $l_{\max} = 2$ and a Gaussian width of 0.4~\AA{} under periodic boundary conditions. The mu1nu1 representation spans elements 1--93 and has 1,116 dimensions. For descriptor $q_{i}$ and the reference set $R_{e}$ with the same central element, the nearest-reference distance is

\begin{equation}
d_{i} = \left[ \max\left( 2 - 2 \max_{\mathbf{q} \in R_{e}} \left( \mathbf{q}_{i} \cdot \mathbf{q} \right), 0 \right) \right]^{1/2}
\label{eq:4}
\end{equation}

Each structure is represented by the 95th percentile of its atomic distances. Three nested pools comprise MPtrj, MPtrj plus sAlex, and MPtrj plus sAlex plus OMat24. The model-resolved comparison uses nine models: DPA-3.1-MPtrj is assigned MPtrj, ORB-v3 and MACE-MPA-0 are assigned MPtrj plus sAlex, and the other plotted models are assigned the full pool. MatterSim-5M and Prophet-OAME-MBD are excluded from this comparison. Distance--error Spearman correlations and median errors within distance quartiles are calculated separately for each model using all 7,599 structures, including errors outside the plotted axis limits.

The atomic error predictor targets the mean Cartesian force error across eleven models. Its 591 initial descriptors encode short interatomic distances, central and neighbouring species, Voronoi geometry, bond valence and neighbouring environments, including bond-orientational parameters $Q_{3}$, $Q_{4}$ and $Q_{6}$\cite{steinhardt1983}. Constant features are removed. Histogram-based gradient boosting\cite{pedregosa2011} fits the logarithmic target with a learning rate of 0.06, 200 iterations, at most 31 leaves, at least 100 samples per leaf and L2 regularization of 5. Five-fold grouped cross-validation keeps related structures and all atoms of each structure together. All 7,599 configurations are included in predictor development. Atomic weights are proportional to $1/N_{j}$ within each sampling source, with source contributions balanced during training. Structure-level test errors use equal structure weights. Atomic predictions are transformed back from logarithms and averaged within each structure.

Sequential comparisons evaluate a constant predictor, a model with four short-distance descriptors and the full model on identical held-out groups. The four descriptors describe minimum, relative minimum and mean distances and the short-contact mass at a relative-distance threshold of 0.9. Coordination and neighbour-environment extensions are separately compared with the base atom model. Paired group bootstrap resampling provides 95\% intervals for prediction-MAE differences. The compression score is one minus the fifth percentile of site-minimum relative distances. Structure force errors are summarized in ten percentile bins of this score and in an $8 \times 8$ percentile map against mean Voronoi heterogeneity. Map cells require at least 20 structures.

\subsection{Geometric heterogeneity and directional distance analysis}

For site $i$, normalized Voronoi-face-area weights $w_{ij}$ define the mean neighbour distance $\bar{r}_{i} = \sum_{j} w_{ij} r_{ij}$. Geometric heterogeneity is the relative weighted mean absolute deviation

\begin{equation}
h_{i} = \frac{\sum_{j} w_{ij} \left| r_{ij} - \bar{r}_{i} \right|}{\bar{r}_{i}}
\label{eq:5}
\end{equation}

Valid tessellations are obtained for 7,583 structures. Site values are shifted by the corresponding same-element median in each training reference. The comparison retains 56 central elements represented in at least five benchmark structures per error group and 100 reference environments. Distributions weight structures equally within each element and then weight elements equally. Five error groups use the mean force error across ten models, excluding Prophet-OAME-MBD.

Periodic neighbour pairs satisfy $r_{ij} \leq 1.25 \max(r_{i,\min}, r_{j,\min})$ and $r_{ij} < 8$~\AA{}, with each undirected pair counted once. Signed offsets $\Delta r = r_{ij} - r_{50,\mathrm{train}}$ distinguish compression and extension relative to the training-reference median for each element pair. The comparison retains 125 pairs supported by all three references and at least five structures in each of five force-error groups, yielding 6,405 contributing structures. Groups are defined by eleven-model mean force MAEs below 60, 60--100, 100--160, 160--250 and at least 250~meV~\AA$^{-1}$. Distance distributions are normalized within each structure and pair, then averaged over structures and equally over pairs. Atomic errors assigned to neighbour pairs use weights proportional to $1/(N_{j} k_{i})$, where $k_{i}$ is the number of selected neighbours. Distance-error curves use 0.2~\AA{} bins supported by at least 20 structures. Compression and extension are matched by element pair and absolute-offset bin, requiring at least five structures on each side and equal weighting of matched bins. Structure bootstrap resampling provides 95\% intervals.

Following UniFFBench\cite{mannan2026}, pairwise interaction curves are calculated with two atoms in a 30~\AA{} cubic cell over separations of 0.10--5.00~\AA{} at 0.05~\AA{} intervals. GRACE's calculator-level minimum-distance guard is disabled for this analysis. The energy minimum within 1--3~\AA{} separates the compressed and extended branches. The shared comparison uses 20 elements with complete curves and interior minima for all eleven models. Mean absolute curvature is obtained from radial-force central differences on each branch. The curves describe model responses without DFT pairwise references. Energy differences, a coarser distance grid and selected 40~\AA{} cell calculations provide numerical checks.

\subsection{Elemental response and angular representation}

Elemental analysis uses the independent Metal-43 dataset and its displacement-response and Fermi-level-response quantities\cite{geng2026}. Force RMSEs are summarized across ten models, excluding Prophet-OAME-MBD. Response measures are standardized across 43 elements, and their standardized sum is standardized again to form the combined response. Benchmark element-specific RMSEs pool Cartesian squared force errors over atoms of each element for each model, then average the resulting RMSEs across models. Spearman correlation compares this profile with the independent elemental difficulty profile.

Geometric descriptors comprise neighbour-distance statistics, bond-angle cosine distributions and volume per atom. DIRECT sampling\cite{qi2024} selects 15,501 configurations from Metal-43 using these descriptors, retaining the original DFT labels. Trajectories are partitioned into contiguous time intervals, with gaps at partition boundaries and duplicate configurations removed. The training, validation and test sets contain 12,782, 1,233 and 1,486 configurations, respectively, and each covers all 43 elements.

Using identical data splits, we compare a radial baseline, spherical-harmonic models with maximum degrees $\ell_{\max} = 2$ and 4, and a parameter-matched radial control for each spherical-harmonic model. Parameter matching is achieved by widening the radial multilayer perceptron that generates tensor-product weights in the interaction block. All configurations share the interaction-layer count, radial-network depth, scalar hidden-channel count, cutoff, radial basis functions and loss function. The radial baseline has 212,288 trainable parameters. The second- and fourth-order comparison groups have 613,952 and 996,352 parameters, respectively. Each configuration is trained for 300 epochs with two random seeds. Element-specific test force RMSEs are averaged over the two runs. Relative RMSE reductions are calculated separately for each element and then averaged with equal element weights.

\section{Data availability}\label{sec:data-availability}

The data underlying this study include benchmark structures, DFT reference labels, model predictions and numerical results supporting the reported analyses. These data are not currently deposited in a public repository.

\section{Code availability}\label{sec:code-availability}

The analysis and plotting code used in this study is not currently deposited in a public repository.


\begin{thebibliography}{40}

\bibitem{yeh2004}
Yeh, J.-W. et al. Nanostructured high-entropy alloys with multiple principal elements: novel alloy design concepts and outcomes. Advanced Engineering Materials 6, 299--303 (2004). \url{https://doi.org/10.1002/adem.200300567}

\bibitem{rost2015}
Rost, C. M. et al. Entropy-stabilized oxides. Nature Communications 6, 8485 (2015). \url{https://doi.org/10.1038/ncomms9485}

\bibitem{vandermaas2023}
van der Maas, E. et al. Investigation of Structure, Ionic Conductivity, and Electrochemical Stability of Halogen Substitution in Solid-State Ion Conductor Li$_3$YBr$_x$Cl$_{6-x}$. The Journal of Physical Chemistry C 127, 125--132 (2023). \url{https://doi.org/10.1021/acs.jpcc.2c07910}

\bibitem{batatia2025}
Batatia, I. et al. A foundation model for atomistic materials chemistry. The Journal of Chemical Physics 163, 184110 (2025). \url{https://doi.org/10.1063/5.0297006}

\bibitem{deng2023}
Deng, B. et al. CHGNet as a pretrained universal neural network potential for charge-informed atomistic modelling. Nature Machine Intelligence 5, 1031--1041 (2023). \url{https://doi.org/10.1038/s42256-023-00716-3}

\bibitem{barrosoluque2026}
Barroso-Luque, L. et al. The Open Materials 2024 (OMat24) inorganic materials dataset and models. Nature Computational Science 6, 642--652 (2026). \url{https://doi.org/10.1038/s43588-026-00996-w}

\bibitem{schmidt2023}
Schmidt, J. et al. Machine-Learning-Assisted Determination of the Global Zero-Temperature Phase Diagram of Materials. Advanced Materials 35, 2210788 (2023). \url{https://doi.org/10.1002/adma.202210788}

\bibitem{zuo2020}
Zuo, Y. et al. Performance and Cost Assessment of Machine Learning Interatomic Potentials. The Journal of Physical Chemistry A 124, 731--745 (2020). \url{https://doi.org/10.1021/acs.jpca.9b08723}

\bibitem{riebesell2025}
Riebesell, J. et al. A framework to evaluate machine learning crystal stability predictions. Nature Machine Intelligence 7, 836--847 (2025). \url{https://doi.org/10.1038/s42256-025-01055-1}

\bibitem{fu2023}
Fu, X. et al. Forces are not Enough: Benchmark and Critical Evaluation for Machine Learning Force Fields with Molecular Simulations. Transactions on Machine Learning Research (2023). \url{https://openreview.net/forum?id=A8pqQipwkt} (preprint: \url{https://doi.org/10.48550/arXiv.2210.07237}).

\bibitem{loew2025}
Loew, A., Sun, D., Wang, H.-C., Botti, S. \& Marques, M. A. L. Universal machine learning interatomic potentials are ready for phonons. npj Computational Materials 11, 178 (2025). \url{https://doi.org/10.1038/s41524-025-01650-1}

\bibitem{du2025}
Du, H. et al. Assessment and Application of Universal Machine Learning Interatomic Potentials in Solid-State Electrolyte Research. ACS Materials Letters 7, 3403--3412 (2025). \url{https://doi.org/10.1021/acsmaterialslett.5c00336}

\bibitem{wines2025}
Wines, D. \& Choudhary, K. CHIPS-FF: Evaluating Universal Machine Learning Force Fields for Material Properties. ACS Materials Letters 7, 2105--2114 (2025). \url{https://doi.org/10.1021/acsmaterialslett.5c00093}

\bibitem{peng2026}
Peng, A. et al. LAMBench: a benchmark for large atomistic models. npj Computational Materials 12, 62 (2026). \url{https://doi.org/10.1038/s41524-025-01929-3}

\bibitem{mannan2026}
Mannan, S. et al. UniFFBench: evaluating universal machine learning force fields against experimental measurements. Nature Computational Science 6, 894--906 (2026). \url{https://doi.org/10.1038/s43588-026-01019-4}

\bibitem{focassio2025}
Focassio, B., Freitas, L. P. M. \& Schleder, G. R. Performance Assessment of Universal Machine Learning Interatomic Potentials: Challenges and Directions for Materials' Surfaces. ACS Applied Materials \& Interfaces 17, 13111--13121 (2025). \url{https://doi.org/10.1021/acsami.4c03815}

\bibitem{deng2025}
Deng, B. et al. Systematic softening in universal machine learning interatomic potentials. npj Computational Materials 11, 9 (2025). \url{https://doi.org/10.1038/s41524-024-01500-6}

\bibitem{tahmasbi2026}
Tahmasbi, H., Knüpfer, A., Kühne, T. D. \& Mirhosseini, H. Benchmarking universal machine learning interatomic potentials on elemental systems. npj Computational Materials 12, 251 (2026). \url{https://doi.org/10.1038/s41524-026-02251-2}

\bibitem{geng2026}
Geng, X., Zhang, W., Wang, L.-W. \& Meng, X. Origin of the machine learning forces field errors across metal elements. npj Computational Materials 12, 102 (2026). \url{https://doi.org/10.1038/s41524-026-01977-3}

\bibitem{rhodes2025}
Rhodes, B. et al. Orb-v3: atomistic simulation at scale. arXiv (2025). \url{https://doi.org/10.48550/arXiv.2504.06231}

\bibitem{lysogorskiy2026}
Lysogorskiy, Y., Bochkarev, A. \& Drautz, R. Graph atomic cluster expansion for foundational machine learning interatomic potentials. npj Computational Materials 12, 114 (2026). \url{https://doi.org/10.1038/s41524-026-01979-1}

\bibitem{kim2026}
Kim, J. et al. Optimizing cross-domain transfer for universal machine learning interatomic potentials. Nature Communications 17, 3432 (2026). \url{https://doi.org/10.1038/s41467-026-70195-8}

\bibitem{kavanagh2026}
Kavanagh, S. R. et al. Fast and Accurate Equivariant Foundation Models for Atomistic Simulation. arXiv (2026). \url{https://doi.org/10.48550/arXiv.2607.28461}

\bibitem{bigi2026}
Bigi, F., Pegolo, P., Mazitov, A., Schmidt, J. \& Ceriotti, M. Pushing the limits of unconstrained machine-learned interatomic potentials. arXiv (2026). \url{https://doi.org/10.48550/arXiv.2601.16195}

\bibitem{xu2026}
Xu, Z., Xie, W. \& Hu, P. Spectral/Spatial Tensor Atomic Cluster Expansion with Universal Embeddings in Cartesian Space. arXiv (2026). \url{https://doi.org/10.48550/arXiv.2509.14961}

\bibitem{zhou2026}
Zhou, Y. et al. MatRIS: Toward Reliable and Efficient Pretrained Machine Learning Interatomic Potentials. arXiv (2026). \url{https://doi.org/10.48550/arXiv.2603.02002}

\bibitem{yang2024}
Yang, H. et al. MatterSim: A Deep Learning Atomistic Model Across Elements, Temperatures and Pressures. arXiv (2024). \url{https://doi.org/10.48550/arXiv.2405.04967}

\bibitem{zhang2026}
Zhang, D. et al. A Graph Neural Network for the Era of Large Atomistic Models. arXiv (2026). \url{https://doi.org/10.48550/arXiv.2506.01686}

\bibitem{kairosmaterials2026}
Kairos Materials. Prophet: atomistic foundation models for materials simulation and discovery. Software and Prophet-OAME-MBD model release (2026). \url{https://github.com/kairosmaterial/prophet}. Model weights: \url{https://huggingface.co/kairosmaterial/prophet}.

\bibitem{matcalccontributors}
matcalc contributors. matcalc: A Python library for calculating materials properties from the potential energy surface. Software. \url{https://github.com/materialyzeai/matcalc} (accessed 7 October 2026).

\bibitem{togo2023}
Togo, A., Chaput, L., Tadano, T. \& Tanaka, I. Implementation strategies in phonopy and phono3py. Journal of Physics: Condensed Matter 35, 353001 (2023). \url{https://doi.org/10.1088/1361-648X/acd831}

\bibitem{ong2013}
Ong, S. P. et al. Python Materials Genomics (pymatgen): A robust, open-source python library for materials analysis. Computational Materials Science 68, 314--319 (2013). \url{https://doi.org/10.1016/j.commatsci.2012.10.028}

\bibitem{rosen}
Rosen, A. quacc -- The Quantum Accelerator. Software. \url{https://doi.org/10.5281/zenodo.7720998} (accessed 7 October 2026).

\bibitem{cao2025}
Cao, G., Yang, S., Ren, J.-C. \& Liu, W. Electronic descriptors for designing high-entropy alloy electrocatalysts by leveraging local chemical environments. Nature Communications 16, 1251 (2025). \url{https://doi.org/10.1038/s41467-025-56421-9}

\bibitem{huang2024}
Huang, Y. et al. Unraveling Reactivity Origin of Oxygen Reduction at High-Entropy Alloy Electrocatalysts with a Computational and Data-Driven Approach. The Journal of Physical Chemistry C 128, 11183--11189 (2024). \url{https://doi.org/10.1021/acs.jpcc.4c01630}

\bibitem{bartok2013}
Bartók, A. P., Kondor, R. \& Csányi, G. On representing chemical environments. Physical Review B 87, 184115 (2013). \url{https://doi.org/10.1103/PhysRevB.87.184115}

\bibitem{himanen2020}
Himanen, L. et al. DScribe: Library of descriptors for machine learning in materials science. Computer Physics Communications 247, 106949 (2020). \url{https://doi.org/10.1016/j.cpc.2019.106949}

\bibitem{steinhardt1983}
Steinhardt, P. J., Nelson, D. R. \& Ronchetti, M. Bond-orientational order in liquids and glasses. Physical Review B 28, 784--805 (1983). \url{https://doi.org/10.1103/PhysRevB.28.784}

\bibitem{pedregosa2011}
Pedregosa, F. et al. Scikit-learn: Machine Learning in Python. Journal of Machine Learning Research 12, 2825--2830 (2011). \url{https://www.jmlr.org/papers/v12/pedregosa11a.html}

\bibitem{qi2024}
Qi, J., Ko, T. W., Wood, B. C., Pham, T. A. \& Ong, S. P. Robust training of machine learning interatomic potentials with dimensionality reduction and stratified sampling. npj Computational Materials 10, 43 (2024). \url{https://doi.org/10.1038/s41524-024-01227-4}

\end{thebibliography}
\end{document}